\documentclass{optica-article}
\journal{opticajournal}
\articletype{Research Article}
\usepackage[section]{placeins}
\usepackage{needspace}
\ifXeTeX
\fi
\renewcommand{\arraystretch}{1.2}
\hypersetup{pdftitle={SGAA: Independent active alignment of optical subgroups with uncompensated aberrations},pdfauthor={Le Dai, Zhenyao Zheng, Wenyong Li, Yuhan Bao, Weijian Hu, Kailun Yang, Kaiwei Wang},pdfsubject={},pdfkeywords={},pdfcreator={LaTeX}}
\begin{document}
\title{SGAA: Independent active alignment of optical subgroups with uncompensated aberrations}
\author{Le Dai, Zhenyao Zheng, Wenyong Li, Yuhan Bao, Weijian Hu, Kailun Yang, and Kaiwei Wang\authormark{*}}
\affil{Zhejiang University}
\email{\authormark{*}daile@zju.edu.cn}
\begin{abstract*}
Uncompensated subgroup aberrations complicate subgroup-level active alignment (AA), as local image-quality extrema do not directly indicate the target alignment state. We propose SGAA, an image-learning-based method for independent active alignment of optical subgroups. Multi-field point spread functions (PSFs) from a dedicated test configuration characterize misalignment, while golden-sample reference calibration and unified regression predict adjustments toward a common optical target across toleranced specimens. In simulations with tolerances, one step reduced the mean differential wavefront RMS from $0.0592\lambda$ to $0.0086\lambda$, an 85.4\% reduction. In alignment experiments on 100 production subgroups, SGAA increased the mean full-lens weighted modulation transfer function (WMTF) at 30, 50, and 80 lp/mm by 0.073, 0.065, and 0.040, respectively, relative to industrial MTF-based scanning. The 8 MP acceptance yield increased by 29.2 percentage points, and mean alignment time decreased from 43 s to 3.1 s. SGAA thus translates independent subgroup reference-state restoration into gains in full-lens imaging, acceptance yield, and alignment speed.
\end{abstract*}
\par\medskip

\section{Introduction}
\label{sec:introduction}

Applications in mobile photography, automotive perception, security monitoring, and ultra-high-definition imaging continue to drive the development of high-resolution imaging systems~\cite{Blahnik2021,Gao2022,Yamashita2013}. In compact high-resolution lenses, tolerance sensitivity to component decenter, tilt, and spacing errors directly affects image-quality consistency and production yield~\cite{Blahnik2021,Ma2013}. Active alignment (AA) uses optical feedback to adjust component poses and guide assembly directly by imaging performance; lens--sensor AA has been used to improve the yield of high-resolution mobile-phone camera modules~\cite{Sahin2017}. On production lines with multiple processes operating in parallel, alignment time affects the number of alignment machines needed to meet production targets. Image quality, alignment cycle time, and investment in equipment are therefore key constraints in method design~\cite{Liu2024,Zhu2026}.

Traditional AA identifies misalignment through aberration models, wavefront measurements, or artificial-star feedback: Gu et al. used nodal aberration theory for the numerical alignment of a three-mirror telescope~\cite{Gu2015}, Heil et al. used artificial-star testing and real-time video holography to guide micro-objective adjustment~\cite{Heil2003}, and Schindlbeck et al. used wavefront measurements for sequential assembly and state estimation of optical systems~\cite{Schindlbeck2018,Schindlbeck2019}. Deep-learning optical metrology learns the mapping between observed images and physical quantities, providing a methodological basis for image-driven parameter estimation~\cite{Zuo2022,LeCun2015,Barbastathis2019}. For complete imaging lenses in mass production, Liu et al. predicted decentering from cross-target images at multiple field points and then performed fine adjustment through a search based on the weighted modulation transfer function (WMTF)~\cite{Liu2024}; Zhu et al. adjusted lens groups and the sensor sequentially using defocused MTF curves, sensitivity correction, and Bayesian optimization~\cite{Zhu2026}. Hu et al. estimated tolerance states from point spread functions (PSFs) at multiple field points and used a surrogate optical model to determine adjustments, whereas Slor et al. diagnosed multi-element misalignments from synthetic spot images; both studies numerically validated image-based learning for state identification~\cite{Hu2025,Slor2026}. At the subsystem level, Lippman and Crowther discussed using an intermediate image to align a subsystem before integrating the remaining parts through a multistep tolerance analysis of a freeform telescope~\cite{Lippman2026}. For local lens-group testing, Zhang et al. used a spatial light modulator and auxiliary optics to complete the imaging function of the test optical path and evaluated subgroup quality using MTF~\cite{Zhang2022}; this use of external optical compensation to enable imaging evaluation is similar to industrial dummy-lens testing.

Multi-group lenses further increase the complexity of optical alignment. A complete lens consists of groups with distinct optical functions, such as the fixed, variator, and compensator groups of a zoom lens, each containing multiple optical elements~\cite{Lee2016}. These groups jointly provide imaging and aberration compensation. Adjusting individual elements simultaneously introduces many coupled degrees of freedom; moving entire groups can reduce the search space~\cite{Zhu2026}, whereas relative errors among components within each group require correction during prealignment. This study addresses how to determine the optical state to be restored by components within a local subgroup and perform their relative active alignment under independent test conditions without the remaining groups of the original lens.

When a local subgroup forms an image independently, the absence of aberration compensation from the remaining groups can produce substantially blurred spots even when the components within the subgroup are in their design states. Extrema of the local MTF or spot size therefore lack a direct correspondence with the target relative state. The two-dimensional spot morphology retains changes caused by misalignment within the subgroup. In the industrial practice on which this study is based, operators prealign components within the subgroup by observing changes in spot morphology and making adjustments by trial and error based on experience. Establishing a mapping from spot morphology to component adjustments and defining a common optical reference for samples with different tolerances are therefore key to automating this step.

\begin{figure}[!htb]
\centering
\includegraphics[width=\linewidth]{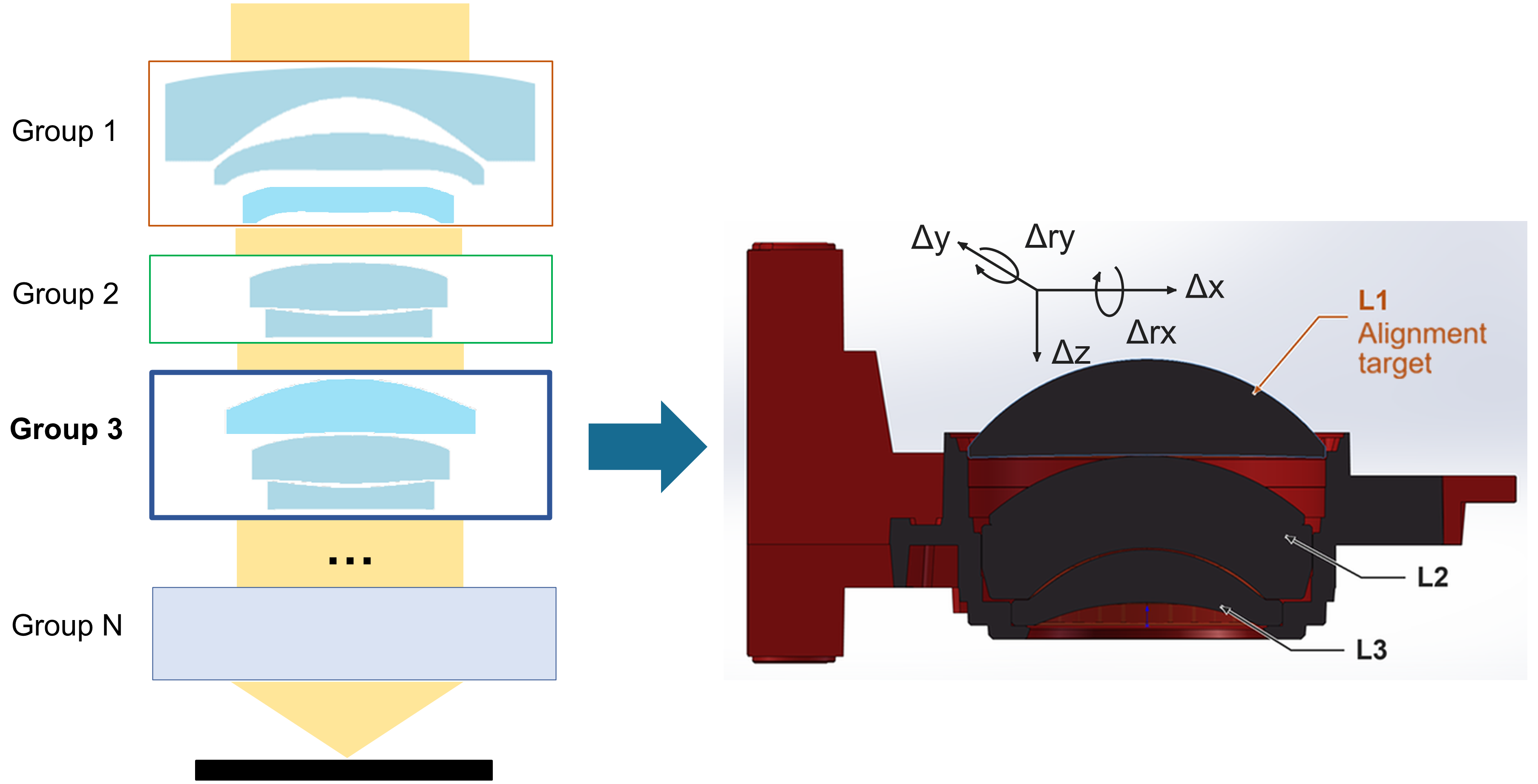}
\caption{Schematic of a divide-and-conquer strategy for group active alignment of a multi-group lens.}
\label{fig:overview}
\end{figure}

As shown in Fig.~\ref{fig:overview}, we propose SGAA, a learning-based method for local optical subgroup alignment that uses independent subgroup prealignment to support stepwise assembly of multi-group lenses. Without incorporating the remaining groups of the original lens, the method acquires PSFs at multiple field points through a dedicated test optical path and learns the mapping from spot morphology to component adjustments. Golden-sample reference calibration establishes a common optical target across samples with different tolerances, and a unified regression network outputs the adjustments toward this reference state.

Restoring the reference state of an independently tested subgroup can improve complete-lens imaging. In five-degree-of-freedom simulations involving 500 samples with tolerances, one SGAA feedback step reduced the mean local defocus-removed differential wavefront RMS from $0.0592\lambda$ to $0.0086\lambda$, a reduction of 85.4\%. The physical AA experiments further showed that SGAA increased the mean complete-lens WMTF at 30, 50, and 80~lp/mm by 0.073, 0.065, and 0.040, respectively, compared with industrial dummy-lens scanning. Using the 8~MP level as the acceptance criterion, the complete-lens yield increased from 52.5\% for the control group to 81.7\%, an increase of 29.2 percentage points, while the mean AA time decreased from 43~s to 3.1~s. These results show that local adjustments under independent subgroup test conditions can improve complete-lens imaging, acceptance yield, and AA cycle time.

The main contributions are as follows.

\begin{enumerate}
\renewcommand{\labelenumi}{(\theenumi)}
\item To the best of our knowledge, this is the first demonstration of active alignment of adjustable components relative to fixed components within a local transmissive subgroup of a mass-produced imaging lens under independent test conditions without the remaining groups of the original lens.

\item We propose a reference calibration method for SGAA that uses the golden sample to estimate specimen-specific reference offsets and correct the label origins, establishing a common optical reference across samples.

\item We use multi-field PSFs to characterize misalignment within the subgroup under non-ideal imaging conditions, replacing manual spot interpretation and trial-and-error adjustment with automatic image-based pose feedback.
\end{enumerate}

Section 2 presents the principles of SGAA. Sections 3 and 4 report the five-degree-of-freedom simulations and two-dimensional physical validation, respectively, and Section 5 summarizes the conclusions and discusses directions for extension.

\section{Methods}\label{sec:methods}

\subsection{Target of local subgroup alignment}\label{sec:methods-target}

Subgroup prealignment requires specifying the optical state that a manufactured subgroup should recover and determining how misalignment within the subgroup can be identified under aberrated imaging conditions.

As shown in Fig.~\ref{fig:overview}, a lens comprising five groups, G1--G5, is used for validation ($N=5$), with G3 as the subgroup under test. L1--L3 denote three assembly components within G3, all retaining manufacturing and assembly tolerances; L1 is adjustable, whereas L2 and L3 are fixed. Let $q$ denote the state of the adjustable component and $q_0$ its target reference state. The misalignment vector is defined as

\begin{equation}
\Delta q=q-q_0=(\Delta x,\Delta y,\Delta r_x,\Delta r_y,\Delta z)^{\mathrm T}.
\label{eq:1}
\end{equation}

Here, $\Delta x$ and $\Delta y$ are lateral decenters, $\Delta r_x$ and $\Delta r_y$ are tilts, and $\Delta z$ is the axial control variable. The simulations cover all five degrees of freedom, whereas the physical experiments validate two-dimensional lateral adjustments, $\Delta x$ and $\Delta y$, of production G3 subgroups. Because local alignment ultimately serves the resolving performance and assembly yield of the complete lens, recovery of the subgroup reference state must be related to complete-lens image quality and acceptance criteria.

\begin{figure}[!htb]
\centering
\includegraphics[width=\linewidth]{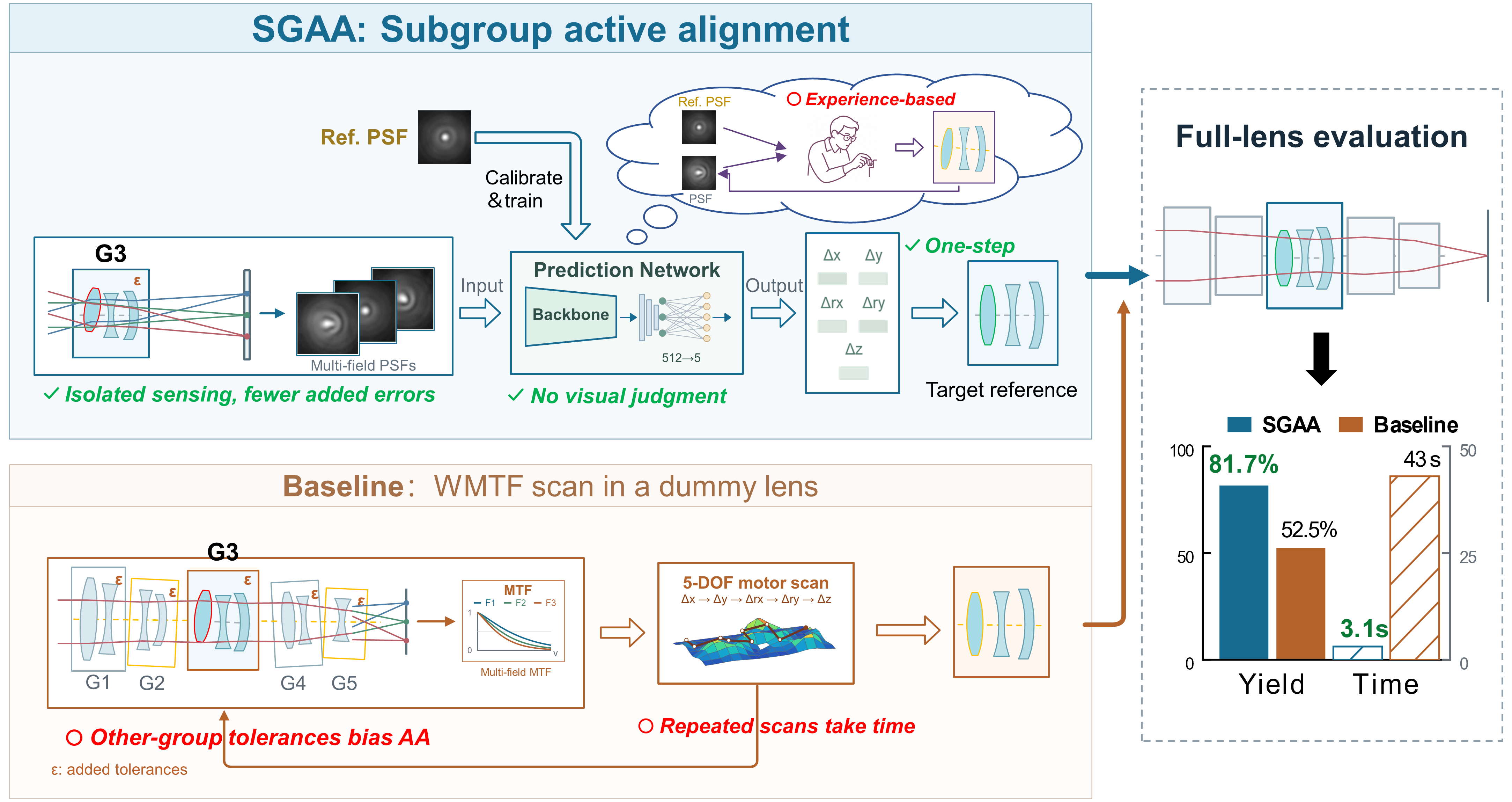}
\caption{Overview of the workflows for SGAA and dummy-lens WMTF scanning. The upper workflow shows SGAA's independent-subgroup PSF observation, reference calibration, and adjustment based on network predictions, motivated by replacing manual interpretation and trial adjustment with a network. The lower workflow uses multi-field WMTF feedback and scanning in a dummy lens. The right-hand side shows the acceptance yield and mean AA time after complete-lens assembly.}\label{fig:2}
\end{figure}

Figure~\ref{fig:2} compares the complete alignment workflows of SGAA and dummy-lens WMTF scanning. Both workflows adjust the movable component within G3 but use different test optical paths and feedback information. The aligned subgroups subsequently undergo a common complete-lens evaluation.

SGAA acquires multi-field PSFs at a fixed image plane in an independent test optical path, without incorporating the other groups of the original lens. Following golden-sample reference calibration and training, the network predicts misalignment, which is converted into actuator corrections. Grayscale PSFs from nine fields near the optical axis are used, with different fields providing complementary responses to the same misalignment. Reference calibration and unified regression are described in Section~\ref{sec:methods-calibration}, and acquisition parameters are given in the experimental settings.

WMTF scanning in a dummy lens is used as the baseline for alignment within the subgroup. Assembly access and process constraints make it difficult to perform subgroup prealignment concurrently with complete-lens assembly. The industrial practice underlying this study therefore uses a fixed dummy lens as the compensating optical path: WMTF scanning adjusts the internal subgroup state, after which the subgroup is removed and installed in a production lens. A previous physical alignment study reported a higher mean WMTF for WMTF search in a complete lens than for direct network regression \cite{Liu2024}, supporting the use of scanning as an accuracy reference.

During scanning, G3 is incorporated into a compensating optical path with tolerances, and candidate states are evaluated using its multi-field WMTF in an axis-by-axis search. After adjustment, the simulation retains G3's tolerances and internal state and places it in a complete lens whose other groups follow the nominal, tolerance-free prescription, enabling common WMTF evaluation with the SGAA results. In the physical experiments, image quality and yield are assessed using the acceptance criteria for the complete lens. Scan budgets, stopping criteria, and timing boundaries are given in Sections 3 and 4 and in the Supplementary Material.

\subsection{PSF observation design under aberrated imaging conditions}\label{sec:methods-observation}

\begin{figure}[!htb]
\centering
\includegraphics[width=\linewidth,height=0.64\textheight,keepaspectratio]{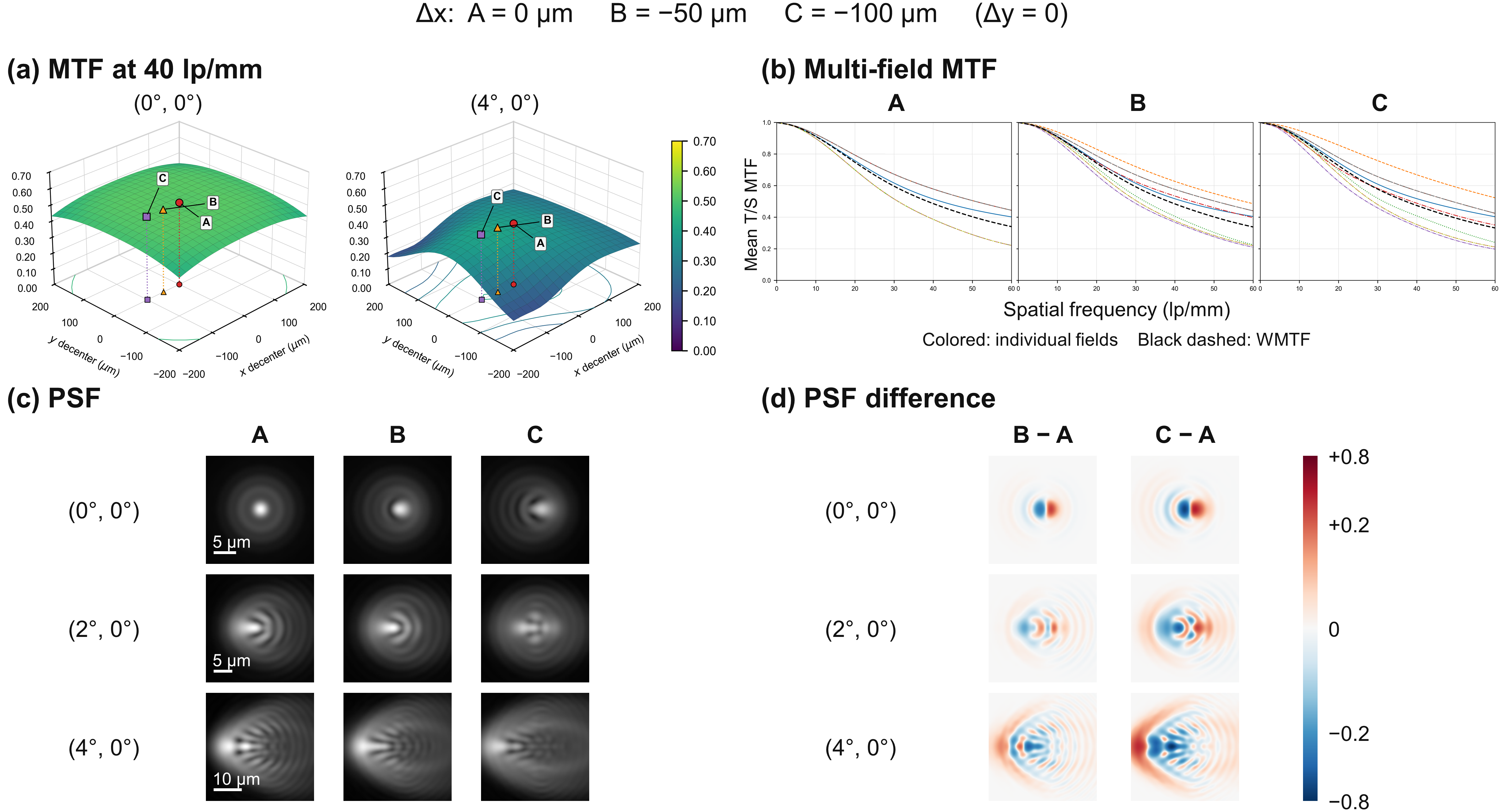}
\caption{Responses of the G3 MTF and PSF to different decenters under independent testing. (a) Distributions of the mean tangential and sagittal MTF at 40 lp/mm versus $\Delta x$ and $\Delta y$ for the central field and the (4$^\circ$, 0$^\circ$) field. (b) Nine-field MTF curves for three states. (c) Polychromatic Huygens PSFs at different fields. (d) Changes in the PSF intensity distribution caused by decenter, with red and blue indicating increases and decreases, respectively.}\label{fig:3}
\end{figure}

During independent testing, a local subgroup lacks aberration compensation from the groups before and after it. Its image-quality metrics therefore no longer correspond directly to the alignment target in the complete lens. Figures~\ref{fig:3}(a) and 3(b) show that, in this example, the subgroup MTF varies relatively weakly or asymmetrically with decenter, and the WMTF curves of several states are similar. Consequently, maximizing the subgroup MTF alone does not directly determine the reference state to which its components should return.

The PSF retains the two-dimensional intensity structure. In Figs.~\ref{fig:3}(c) and 3(d), decenter changes the main lobe, sidelobes, and energy distribution, while the difference maps reveal the locations and directions of these changes. Following the industrial practice of judging misalignment from spot patterns, SGAA learns a mapping between these morphological changes and component pose.

For the same sample with fixed manufacturing tolerances, at a fixed field $h$ and wavelength $\lambda$, an equivalent-pupil imaging representation gives the complex image-plane amplitude and PSF as~\cite{Fienup1993,Fienup1982}

\begin{equation}
U_h(q)=\mathcal{F}_h\!\left\{A_h(q)e^{ikW_h(q)}\right\},\qquad
I_h(q)=|U_h(q)|^2,\qquad k=2\pi/\lambda.
\label{eq:2}
\end{equation}

Here, $A_h$ and $W_h$ are the pupil amplitude and wavefront, respectively, and $\mathcal{F}_h$ denotes the Fourier imaging relation at the fixed image plane. The subscript $r$ denotes the target reference state of the same sample. Defining $\delta A_h=A_h-A_{h,r}$ and $\delta W_h=W_h-W_{h,r}$, the change in the image-plane field is

\begin{equation}
\begin{aligned}
\delta U_h&=\mathcal{F}_h\!\left\{e^{ikW_{h,r}}\left[(A_{h,r}+\delta A_h)e^{ik\delta W_h}-A_{h,r}\right]\right\}\\
&\simeq\mathcal{F}_h\!\left\{e^{ikW_{h,r}}(\delta A_h+ikA_{h,r}\delta W_h)\right\}.
\end{aligned}
\label{eq:3}
\end{equation}

The second line of Eq. \eqref{eq:3} requires only a small additional phase perturbation $|k\delta W_h|\ll 1$ and a small amplitude change; the reference wavefront may contain substantial intrinsic aberrations. The corresponding intensity difference is

\begin{equation}
\delta I_h=2\operatorname{Re}(U_{h,r}^{*}\delta U_h)+|\delta U_h|^2.
\label{eq:4}
\end{equation}

Observation sensitivity describes the extent to which the optical response changes under a small component adjustment~\cite{Melchert2022} and underlies misalignment identification from PSFs. Equations \eqref{eq:3}--\eqref{eq:4} show that the adjustment-induced field change and the reference field jointly produce an observable intensity response. PSF changes can therefore retain misalignment information even in the presence of substantial intrinsic aberration. SGAA accordingly learns the relationship between multi-field spot patterns and known adjustment amounts, extracting pose features from observations in which intrinsic aberration and alignment response are coupled. Different fields provide complementary responses, and their joint use supplies more information for misalignment identification~\cite{Wang2026Field}. The propagation approximation, polychromatic observations, and applicability conditions are discussed in Section S1 of the Supplementary Material.

Figure~\ref{fig:4} establishes the PSF observation chain from the point source to the detector. Both the G3 simulations and real-data acquisition use an infinite conjugate. The subgroup first forms a spot at an intermediate image plane, converting adjustment-induced changes in the outgoing field into an intensity structure. A microscope objective and tube lens then magnify this intermediate image onto the detector to improve sampling of spot details. The two imaging stages respectively form the misalignment response and magnify it for readout. For other subgroups, the object distance can be selected according to their focal lengths.

\begin{figure}[!htb]
\centering
\includegraphics[width=\linewidth,height=0.64\textheight,keepaspectratio]{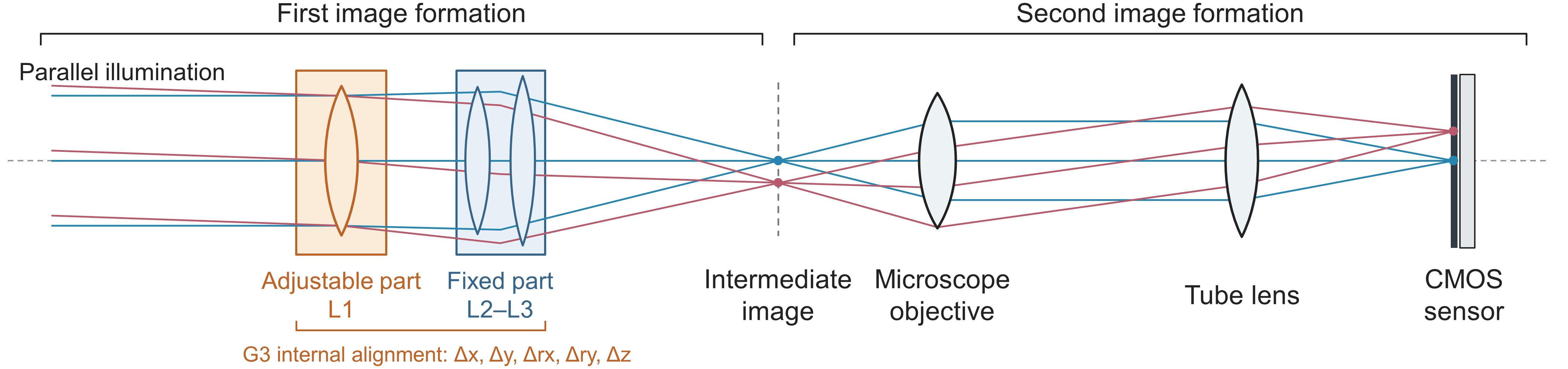}
\caption{Independent-subgroup PSF observation optical path.}\label{fig:4}
\end{figure}

\subsection{Learning and alignment guided by a golden-sample reference}\label{sec:methods-calibration}

\begin{figure}[!htb]
\centering
\includegraphics[width=\linewidth,height=0.64\textheight,keepaspectratio]{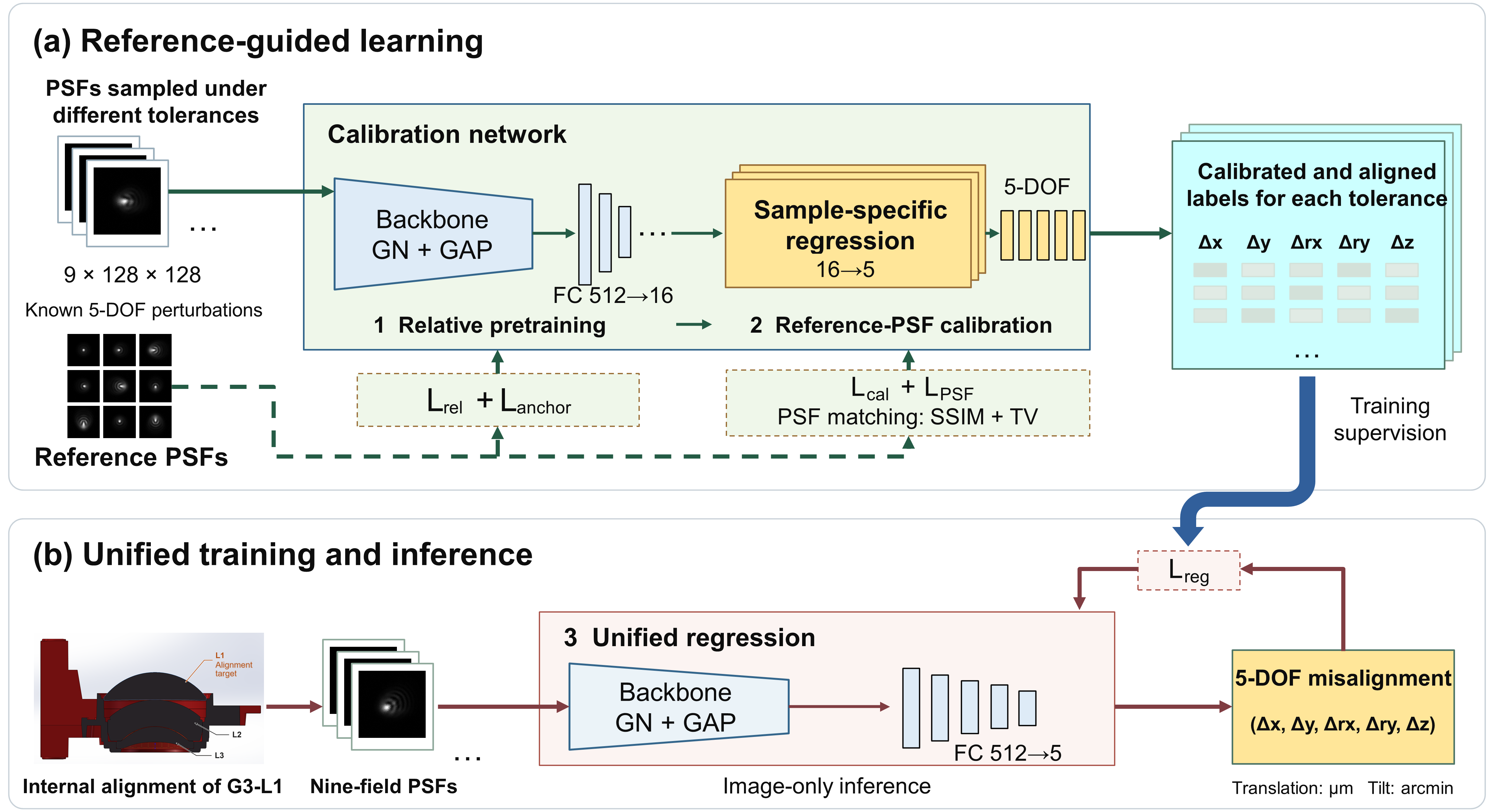}
\caption{Network framework guided by a golden-sample reference. (a) A shared feature extractor and regression branches first undergo relative pretraining, after which acquisition coordinates and reference PSFs are used to calibrate the reference positions and generate corrected labels. (b) A unified regression network is trained using the corrected labels and predicts misalignment from multi-field PSFs during deployment. GN, GAP, and FC denote group normalization, global average pooling, and fully connected layers, respectively. SSIM and TV assess PSF structural similarity and differences in normalized energy distributions, respectively.}\label{fig:5}
\end{figure}

Manufacturing and assembly tolerances alter the intrinsic aberration of each sample~\cite{Zhou2024}, so the zero of the acquisition coordinates need not correspond to a common target optical state. The alignment reference must therefore be calibrated for each sample alongside learning the response to alignment motion. A G3 with tolerances, prealigned by complete-lens MTF scanning, is selected as the golden sample. A small set of its reference PSFs, including perturbation and through-focus images, provides optical constraints. This reference calibration is performed offline; PSFs are acquired at a fixed image plane during deployment. The network framework in Fig.~\ref{fig:5} sequentially performs relative learning, reference calibration, and unified regression.

Under identical test conditions and with the same reference sphere, omitting the field and pupil coordinates, let $W_0$ be the design intrinsic wavefront of the subgroup without tolerances, $q_i^r$ the calibrated reference state of sample $i$, $B_i=W_i(q_i^r)-W_0$ the reference residual, and $D_i(q)=W_i(q)-W_i(q_i^r)$ the motion-induced change. After removal of the common reference-sphere phase, the complex field is

\begin{equation}
E_i^{\mathrm{red}}(q)=A_i(q)\,
\underbrace{e^{ikW_0}}_{\text{Design aberration}}\,
\underbrace{e^{ikB_i}}_{\text{Reference residual}}\,
\underbrace{e^{ikD_i(q)}}_{\text{Motion-induced change}}.
\label{eq:5}
\end{equation}

$W_0$ may contain substantial intrinsic aberration. $B_i$ retains differences arising jointly from tolerances and reference adjustment, while the response of $D_i$ is also affected by tolerances. For the golden sample at its reference state, $D_g(q_g^r)=0$, but $B_g$ need not be zero. The wavefront difference between an ordinary sample and the golden sample is

\begin{equation}
W_i(q)-W_g(q_g^r)=\underbrace{B_i-B_g}_{\text{Reference difference}}
+\underbrace{D_i(q)}_{\text{Motion-induced change}}.
\label{eq:6}
\end{equation}

Equations \eqref{eq:5}--\eqref{eq:6} distinguish the design intrinsic aberration, sample-to-sample reference differences, and adjustment-induced changes, which jointly form the PSF observations. Using known differences in adjustment, relative pretraining enables the shared feature extractor $E_\theta$ and sample-specific regression branch $h_i$ to learn the response to alignment motion and extract features for pose estimation. Response learning and reference calibration serve different purposes: relative supervision relates observation changes to adjustment amounts, whereas subsequent reference calibration determines the optical state to which this response should return. Loss definitions are given in Section S2 of the Supplementary Material.

Offline reference calibration relates the response to alignment motion to the target state represented by the golden sample. While fitting the acquisition coordinates, all branches use the same golden-sample reference PSFs to establish a common optical reference and constrain the PSF response near the estimated reference position. Corrected labels are thereby constructed for each sample using a common optical target. Specifically, the golden-sample images are passed through each branch, and the reference offset is estimated as the mean difference between the predictions and the known golden-sample labels. The supervision targets are then defined as

\begin{equation}
\begin{aligned}
\widehat{O}_i&=\frac{1}{N_g}\sum_{g=1}^{N_g}\left[h_i(E_\theta(y_g))-u_g^{\mathrm{ref}}\right],\\
\widetilde{u}_{ij}&=u_{ij}-\widehat{O}_i.
\end{aligned}
\label{eq:7}
\end{equation}

Here, $N_g$ is the number of golden-sample reference images; $y_g$ and $u_g^{\mathrm{ref}}$ are the input and known label, respectively, of the $g$th golden-sample image set, with all known labels equal to zero. $y_{ij}$ and $u_{ij}$ denote the PSF input and acquisition label, respectively, of sample $i$ at its $j$th state. $\widehat{O}_i$ is the corresponding estimated reference offset, and $\widetilde{u}_{ij}$ is the corrected label representing sample misalignment relative to the estimated common reference. The PSF reference constraint is evaluated over a sampled neighborhood around the estimated position, as detailed in Section S2 of the Supplementary Material.

After reference calibration, the corrected labels are used to train the unified regression network $F_\varphi$ in Fig.~\ref{fig:5}(b):

\begin{equation}
\begin{aligned}
L_{\mathrm{reg}}&=\mathbb{E}_{i,j}\,\ell_{\mathrm{abs}}\!\left(F_\varphi(y_{ij}),\widetilde{u}_{ij}\right),\\
\widehat{\Delta q}&=F_\varphi(y).
\end{aligned}
\label{eq:8}
\end{equation}

Here, $\ell_{\mathrm{abs}}$ is the coordinate regression loss. Using the corrected labels, the unified network learns misalignment relative to the reference state without directly solving for the individual wavefront terms. During deployment, multi-field PSFs are acquired at a single fixed image plane and fed into the unified network to obtain a misalignment prediction, which is converted into actuator corrections.

The SGAA prediction ultimately relates the subgroup imaging state to complete-lens imaging. With the other optical paths in the complete lens fixed, let $q_i^r$ denote the calibrated reference state, $K_{i,h}^{\mathrm{full}}$ the complete-lens propagation kernel including propagation and aperture effects, and $E_{\mathrm{in},h}$ the fixed incident field. $\rho$ and $P$ are the coordinates at the input plane and common exit-pupil plane, respectively. The field and wavefront differences between the current and reference states are

\begin{equation}
\begin{aligned}
\delta E_{i,h}^{\mathrm{full}}(P;q)&=\int\!\bigl[K_{i,h}^{\mathrm{full}}(P,\rho;q)-K_{i,h}^{\mathrm{full}}(P,\rho;q_i^r)\bigr]E_{\mathrm{in},h}(\rho)\,\mathrm{d}^{2}\rho,\\
\delta W_{i,h}^{\mathrm{full}}(P;q)&=\frac{1}{k}\operatorname{Arg}_{\mathrm{cont}}\!\left[E_{i,h}^{\mathrm{full}}(P;q)E_{i,h}^{\mathrm{full}}(P;q_i^r)^{*}\right]\to 0\quad(q\to q_i^r).
\end{aligned}
\label{eq:9}
\end{equation}

Within a neighborhood where propagation varies smoothly with state and no abrupt obscuration occurs, the field difference tends to zero as the component returns to its reference state. When the reference field is nonzero and its phase can be unwrapped continuously, the wavefront difference also tends to zero. By restoring the component reference state, SGAA can therefore recover its reference optical contribution within the complete lens and its aberration-compensation relationship with the other groups. This explains how local subgroup prealignment can improve complete-lens imaging. Relevant propagation conditions are given in Section S1 of the Supplementary Material, and the practical AA effect is examined by simulation in Section 3.

\section{Simulation study}\label{sec:simulation}

This section applies the SGAA method presented in Section~2 to five-degree-of-freedom active alignment of L1 within the toleranced G3 subgroup. We first compare the subgroup states and complete-lens image quality obtained with the scanning baselines and SGAA, and then examine the role of the reference calibration scheme. Local wavefronts are evaluated as differences from the nominal, tolerance-free subgroup wavefront, while complete-lens image quality is evaluated using the equally weighted mean MTF over multiple fields.

\subsection{Simulation settings}\label{sec:simulation-settings}

\subsubsection{Optical configuration and sampling}\label{sec:simulation-optical}

The simulation considers G3 in Fig.~1, with L1 adjustable and L2 and L3 fixed. The manufacturing, internal component assembly, and whole-group positioning tolerances listed in Table~\ref{tab:tolerances} were set according to practical production conditions. Different specimens were generated by Monte Carlo sampling in Zemax. Manufacturing and internal assembly tolerances were sampled from zero-mean normal distributions truncated at $\pm$2$\sigma$. Sampling also included a fixed initial mechanical offset of L1 and whole-group placement errors of G3. The golden sample also contained tolerances.

\begin{table}[!htb]
\centering
\caption{Main manufacturing and assembly tolerances and sampling-center positioning errors}
\label{tab:tolerances}
\fontsize{8}{10}\selectfont
\setlength{\tabcolsep}{2.5pt}
\begin{tabular}{@{}l@{\hspace{4pt}}r@{\hspace{22pt}}l@{\hspace{4pt}}r@{}}
\cline{1-2}\cline{3-4}
\multicolumn{2}{c}{Manufacturing tolerances} & \multicolumn{2}{c}{Assembly and positioning} \\
Parameter & Range & Parameter & Range \\
\cline{1-2}\cline{3-4}
Surface decenter ($x$, $y$) & $\pm5~\mu$m & Component decenter ($x$, $y$) & $\pm20~\mu$m \\
Center thickness & $\pm10~\mu$m & Component tilt ($r_x$, $r_y$) & $\pm4$ arcmin \\
Fixed air spacing & $\pm10~\mu$m & Group decenter ($x$, $y$) & $\pm20~\mu$m \\
Surface tilt ($r_x$, $r_y$) & $\pm2$ arcmin & Group tilt ($r_x$, $r_y$) & $\pm4$ arcmin \\
Surface power error & $\pm4$ fringe & Group axial position / gap & $\pm20~\mu$m \\
Surface irregularity & $\pm0.4$ fringe & Sampling-center decenter ($x$, $y$) & $\pm30~\mu$m \\
Refractive-index error & $\pm0.0003$ & Sampling-center tilt ($r_x$, $r_y$) & $\pm12$ arcmin \\
Relative Abbe-number error & $\pm0.3\%$ & Sampling-center axial position / gap & $\pm30~\mu$m \\
\cline{1-2}\cline{3-4}
\end{tabular}
\par\smallskip
\begin{minipage}{\linewidth}\footnotesize
Note: The reference wavelength for surface figure errors was 550 nm.\par
\end{minipage}
\end{table}

Local observations used collimated illumination and a fixed image plane. Polychromatic Huygens PSFs were calculated at nine fields: the center and $\pm$1$^\circ$ and $\pm$2$^\circ$ along the X and Y directions. For each specimen, 1024 Latin hypercube samples~\cite{McKay1979} were taken over $\pm$60 $\mu$m for each of $\Delta x$ and $\Delta y$, $\pm$20 arcmin for each of $\Delta r_x$ and $\Delta r_y$, and $-$45 to +75 $\mu$m for $\Delta z$. The pixel size was 3 $\mu$m. The nine 128 $\times$ 128 grayscale PSFs were individually min--max normalized and stacked in a fixed order. Gaussian noise at a signal-to-noise ratio of 28 dB and subpixel perturbations with a standard deviation of 0.4 pixel were added to the training inputs \cite{Hu2025}.

The scanning comparisons used a toleranced compensating optical path (dummy lens) and a tolerance-free compensating optical path (Perfect lens). The compensating groups of the former included manufacturing and whole-group positioning errors, whereas those of the latter used the nominal, tolerance-free prescription. Both retained the tolerances of the G3 subgroup being adjusted.

After AA with each method, the internal tolerances of G3 and the resulting L1 state were retained. All results were evaluated under common complete-lens conditions, with every other group using its nominal, tolerance-free prescription, to evaluate image quality after G3 alignment. The evaluation fields were the center, ($\pm$20$^\circ$, 0$^\circ$), and (0$^\circ$, $\pm$20$^\circ$). The tangential and sagittal MTFs were averaged at each field, followed by an equally weighted average over the five fields \cite{Liu2024}. The value at 60 lp/mm is denoted WMTF60.

\subsubsection{Network and training settings}\label{sec:simulation-training}

The simulation used ResNet18~\cite{He2016} with GroupNorm~\cite{Wu2018}, taking nine-channel PSFs as input and predicting five degrees of freedom. Following relative pretraining and golden-sample reference calibration according to Section~2, the corrected labels were used to train the unified regression network, whose backbone was initialized independently with ImageNet-pretrained weights~\cite{Russakovsky2015}. The degrees of freedom were normalized by their physical scales to form the loss, and AdamW~\cite{Kingma2015,Loshchilov2019} was used for optimization. Details of the losses and calibration are provided in Supplement 1, Section S2.

The training, validation, independent test, and AA evaluation sets contained 22, 10, 18, and 500 subgroups, respectively. The four sets did not overlap and used the same optical parameters and tolerance allocations. The training and validation sets were used for relative pretraining and reference calibration. The unified regressor was trained on the training set, and the validation set was used for model selection. The independent test set was used to evaluate five-degree-of-freedom prediction performance, while the AA evaluation set was used to compare alignment endpoints and reference calibration schemes.

\subsection{Simulation experiments}\label{sec:simulation-experiments}

\subsubsection{Evaluation of subgroup states after AA}\label{sec:simulation-state}

Scanning used the dummy lens and Perfect lens defined in Section~\ref{sec:simulation-optical}, adjusting the five degrees of freedom of L1 sequentially with the WMTF of the corresponding optical configuration as feedback. Two endpoints were compared: a budget of 50 MTF scans and a complete scan with no preset limit on the number of MTF scans (full). The detailed procedure is provided in the Supplementary Material.

For each of the 500 specimens, all methods started from the same initial misalignment state. The mean decenter magnitudes after one SGAA step and complete scans with the dummy lens and Perfect lens were 12.84, 44.32, and 45.32 $\mu$m, respectively. The corresponding mean tilt magnitudes were 3.49, 14.67, and 15.39 arcmin. Smaller magnitudes indicate subgroup states closer to the design geometric center.

To assess whether the subgroup states approached the design reference wavefront, local wavefronts before and after one SGAA feedback step were compared in pairs for the 500 toleranced specimens. Wavefront differences were calculated relative to the nominal, tolerance-free G3 wavefront at the central field, 550 nm, and the fixed local nominal image plane. Piston, tilt, and defocus were removed analytically. A smaller differential wavefront indicates a subgroup state closer to the design reference.

\begin{figure}[!htb]

\centering

\includegraphics[width=\linewidth,height=0.50\textheight,keepaspectratio]{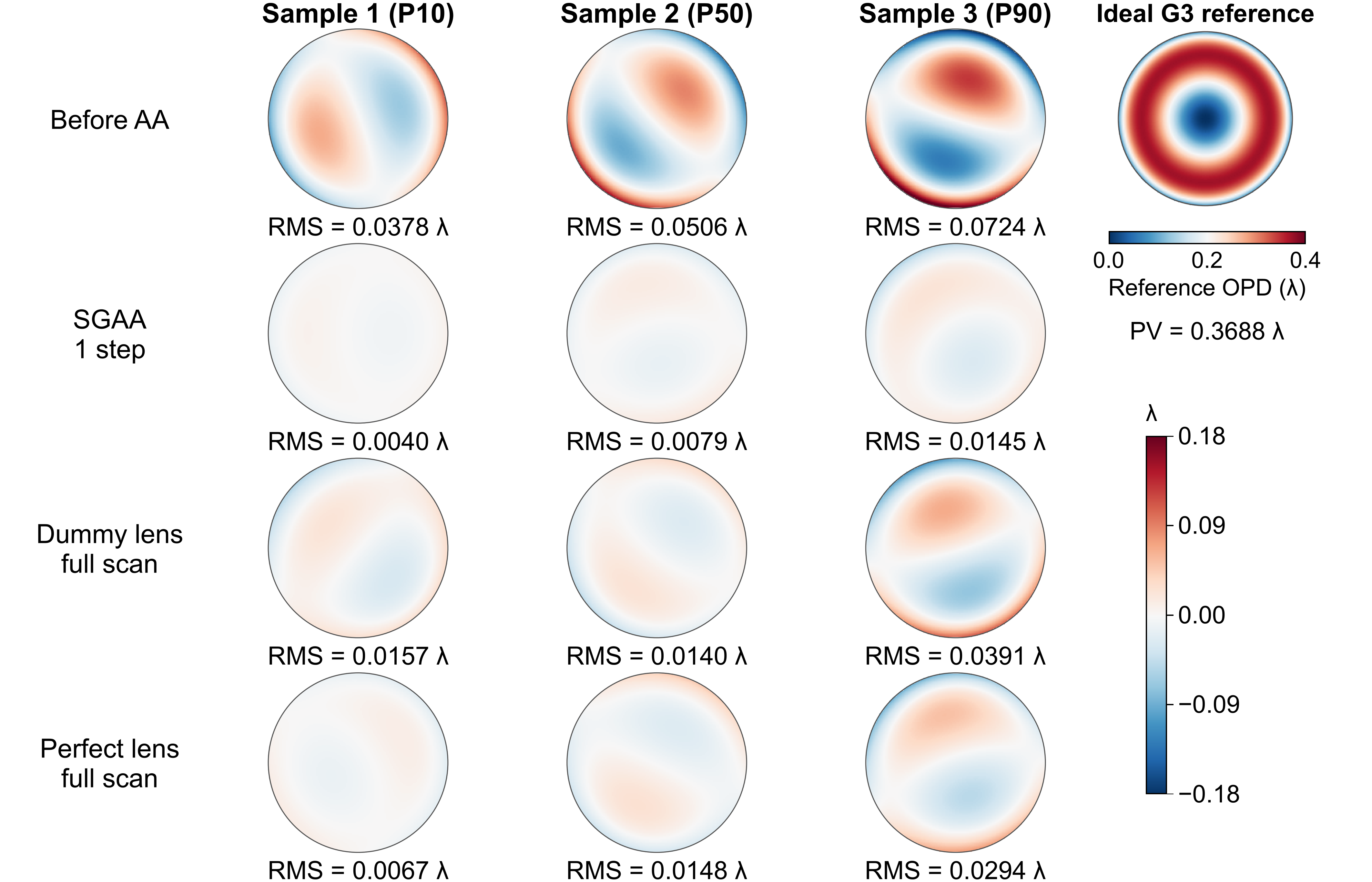}

\caption{Defocus-removed differential wavefronts of three representative G3 specimens before alignment, after one SGAA feedback step, and after the two types of complete-lens scans. The three columns correspond to specimens at the 10th, 50th, and 90th percentiles of the differential RMS after SGAA feedback. The upper-right panel shows the nominal, tolerance-free G3 reference wavefront subtracted in the calculation.}

\label{fig:simulation-wavefront}

\end{figure}

\begin{table}[!htb]
\centering
\caption{Local differential wavefront statistics in different alignment states (unit: $\lambda$)}
\label{tab:wavefront}
\fontsize{8}{10}\selectfont
\setlength{\tabcolsep}{4pt}
\begin{tabular}{@{}lcccc@{}}
\hline
Metric (λ) & Before AA & \shortstack{SGAA\\1 step} & \shortstack{Dummy lens\\full scan} & \shortstack{Perfect lens\\full scan} \\
\hline
Differential RMS, defocus removed & 0.0592 $\pm$ 0.0275 & {\bfseries 0.0086 ± 0.0041} & 0.0158 $\pm$ 0.0151 & 0.0116 $\pm$ 0.0067 \\
Differential PV, defocus removed & 0.3366 $\pm$ 0.1567 & {\bfseries 0.0487 ± 0.0238} & 0.0900 $\pm$ 0.0863 & 0.0668 $\pm$ 0.0383 \\
Coma amplitude & 0.0590 $\pm$ 0.0276 & {\bfseries 0.0080 ± 0.0043} & 0.0152 $\pm$ 0.0153 & 0.0111 $\pm$ 0.0067 \\
Astigmatism amplitude & 0.0032 $\pm$ 0.0021 & {\bfseries 0.0012 ± 0.0007} & 0.0022 $\pm$ 0.0016 & 0.0023 $\pm$ 0.0016 \\
Spherical amplitude & 0.0032 $\pm$ 0.0021 & 0.0023 $\pm$ 0.0016 & 0.0025 $\pm$ 0.0018 & {\bfseries 0.0019 ± 0.0015} \\
Defocus amplitude & 0.0648 $\pm$ 0.0480 & {\bfseries 0.0520 ± 0.0380} & 0.0567 $\pm$ 0.0449 & 0.0520 $\pm$ 0.0396 \\
\hline
\end{tabular}
\par\smallskip
\begin{minipage}{\linewidth}\footnotesize
Note: Piston and tilt were removed for RMS/PV. Aberration amplitudes were calculated from differences in Zemax Standard coefficients.\par
\end{minipage}
\end{table}

\FloatBarrier

Figure~\ref{fig:simulation-wavefront} and Table~\ref{tab:wavefront} show that one SGAA feedback step reduced the mean defocus-removed differential RMS of the 500 specimens from 0.0592 $\lambda$ to 0.0086 $\lambda$, an 85.4\% reduction, outperforming the value of 0.0116 $\lambda$ obtained with the Perfect lens complete scan. In paired comparisons, 324/500 specimens (64.8\%) achieved a lower differential RMS. The three representative specimens shown in Fig.~\ref{fig:simulation-wavefront} also had lower differential RMS values than the endpoints of both complete scans, demonstrating effective recovery of subgroup wavefronts close to the design reference. Coma was the dominant differential aberration in this batch. SGAA reduced the mean coma amplitude from 0.0590 $\lambda$ to 0.0080 $\lambda$, an 86.5\% reduction, below the Perfect lens complete-scan value of 0.0111 $\lambda$. This demonstrates effective correction of the dominant asymmetric aberration.

\subsubsection{Image quality in the complete lens}\label{sec:simulation-image}

For the 500 alignment specimens, complete-lens image quality after assembly was compared for the subgroup states obtained from one SGAA feedback step, the L1 design geometric center (Nominal), and scanning, starting from the same initial states. Nominal sets all five degrees of freedom of L1 to zero, without additional compensation for aberrations caused by the internal tolerances of G3. The subgroup states obtained with all methods were evaluated under the common complete-lens conditions defined in Section~\ref{sec:simulation-optical}. The tolerance-free complete-lens design has a WMTF60 of 0.7484.

\begin{table}[!htb]
\centering
\caption{Complete-lens MTF60 in different alignment states}
\label{tab:mtf}
\fontsize{8}{10}\selectfont
\setlength{\tabcolsep}{3pt}
\begin{tabular}{@{}lcc@{}}
\hline
State & Center-field MTF60 & Edge-field mean MTF60 \\
\hline
\textit{Initial} & 0.1388 $\pm$ 0.0808 & 0.1326 $\pm$ 0.0697 \\
\textit{Nominal} & 0.3888 $\pm$ 0.1640 & 0.3394 $\pm$ 0.1319 \\
{\bfseries\itshape SGAA, 1 step} & {\bfseries 0.6012 ± 0.0896} & {\bfseries 0.5502 ± 0.0715} \\
\textit{Dummy lens, 50} & 0.3187 $\pm$ 0.1106 & 0.3316 $\pm$ 0.1109 \\
\textit{Dummy lens, full} & 0.4007 $\pm$ 0.0933 & 0.4089 $\pm$ 0.1127 \\
\textit{Perfect lens, 50} & 0.5000 $\pm$ 0.1333 & 0.4494 $\pm$ 0.1119 \\
\textit{Perfect lens, full} & {\bfseries 0.6870 ± 0.1138} & {\bfseries 0.5808 ± 0.0967} \\
\hline
\end{tabular}
\par\smallskip
\begin{minipage}{\linewidth}\footnotesize
Note: The labels \textit{50} and \textit{full} denote, respectively, a budget of 50 MTF scans and a complete scan with no preset limit on the number of MTF scans. \textit{Perfect lens, full} averaged 152 scans.\par
\end{minipage}
\end{table}

\begin{figure}[!htb]

\centering

\includegraphics[width=\linewidth]{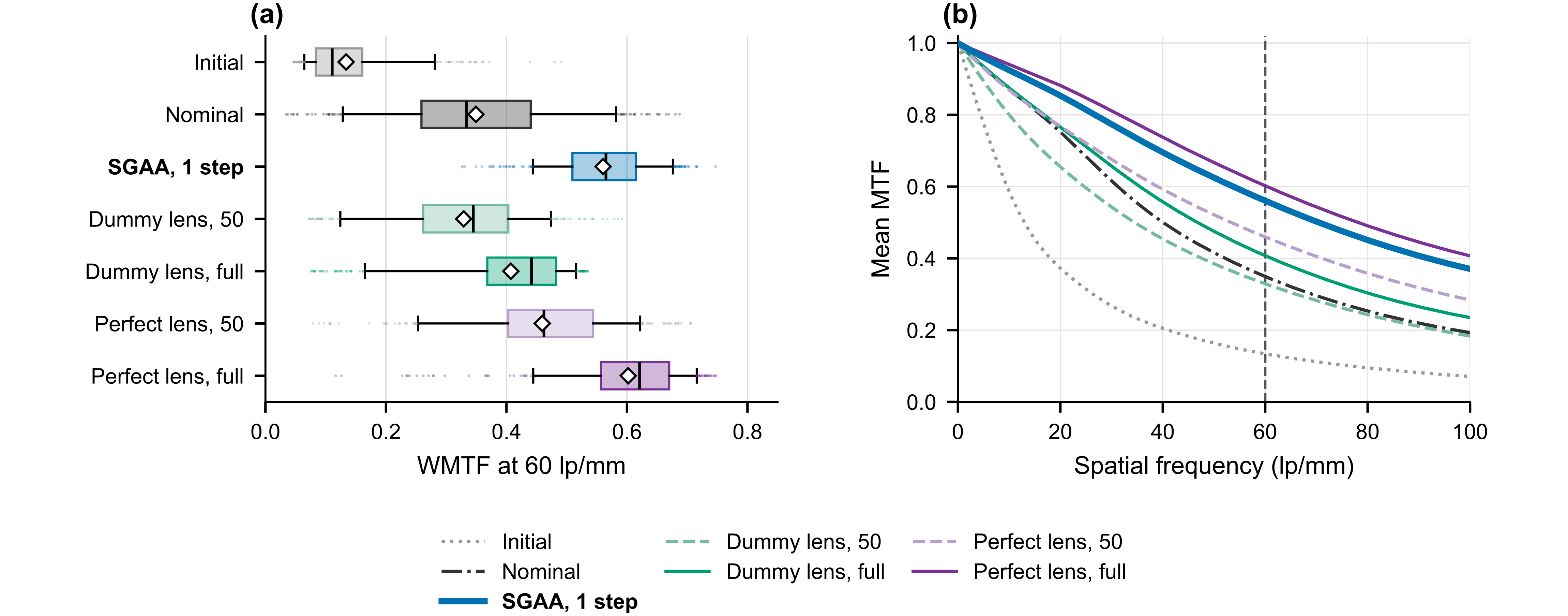}

\caption{Complete-lens image quality of the 500 specimens in different alignment states. (a) Distributions of equally weighted five-field WMTF60. (b) MTF curves averaged over the five fields and 500 specimens. The boxes and dashed lines indicate the 25th--75th and 5th--95th percentiles, respectively; the lines within the boxes and diamonds indicate the medians and means, respectively.}

\label{fig:simulation-mtf}

\end{figure}

Figure~\ref{fig:simulation-mtf} and Table~\ref{tab:mtf} show that, under common complete-lens evaluation conditions, one SGAA feedback step increased WMTF60 for the 500 toleranced specimens from 0.1338 $\pm$ 0.0710 to 0.5604 $\pm$ 0.0735, exceeding the Nominal value of 0.3493 obtained by directly returning to the design geometric center. This indicates that reference-guided adjustment improves complete-lens imaging. With one prediction and adjustment, SGAA achieved a mean WMTF60 exceeding the dummy lens and Perfect lens results under a 50-scan budget by 0.2313 and 0.1009, respectively, and the toleranced dummy lens complete-scan result by 0.1531. These results demonstrate the marked advantages of independent subgroup pre-alignment under limited measurement budgets and practical compensating-optics tolerances. Averaging 152 scans, \textit{Perfect lens, full} provides an idealized upper benchmark by excluding compensating-optics tolerances and implementation-related degradation. Practical compensating optics inevitably contain tolerances and cannot fully meet these conditions. Even when referenced to \textit{Perfect lens, full}, one SGAA feedback step reached 93.1\% of its mean WMTF60 (0.5604 versus 0.6020). The central-field and four-edge-field mean MTF60 values after one step were 0.6012 $\pm$ 0.0896 and 0.5502 $\pm$ 0.0715, respectively, showing improvement across central and edge fields.

Together, the reductions in local differential wavefronts and the improvements in complete-lens WMTF show that local correction based on independently observed subgroup PSFs can improve both the subgroup state and complete-lens imaging after assembly.

\subsection{Comparison of reference calibration schemes}\label{sec:simulation-calibration}

To evaluate the effect of reference calibration on unified regression learning and alignment, four schemes were compared: SGAA learned reference calibration, PSF second-moment calibration, PSF-similarity calibration, and an uncalibrated scheme. The first three established the reference using golden-sample PSFs and trained a unified regression network after offline calibration. The uncalibrated scheme trained a regressor directly on mechanical motor coordinates.

SGAA used the method described in Section~2. The second-moment scheme fitted a five-dimensional quadratic model to the mean central second moments of the nine-field PSFs, estimated bounded reference positions for the specimen and golden sample separately, and corrected the labels using their difference. The PSF-similarity scheme used the golden-sample PSFs as templates and selected the state with the smallest composite PSF difference (SSIM~\cite{Wang2004SSIM}) among each specimen's acquired states as the reference. Complete-lens image quality for all schemes was evaluated using the same 500 specimens as in Section~\ref{sec:simulation-experiments}, with one prediction and adjustment step from common initial states.

\begin{figure}[!htb]

\centering

\includegraphics[width=\linewidth]{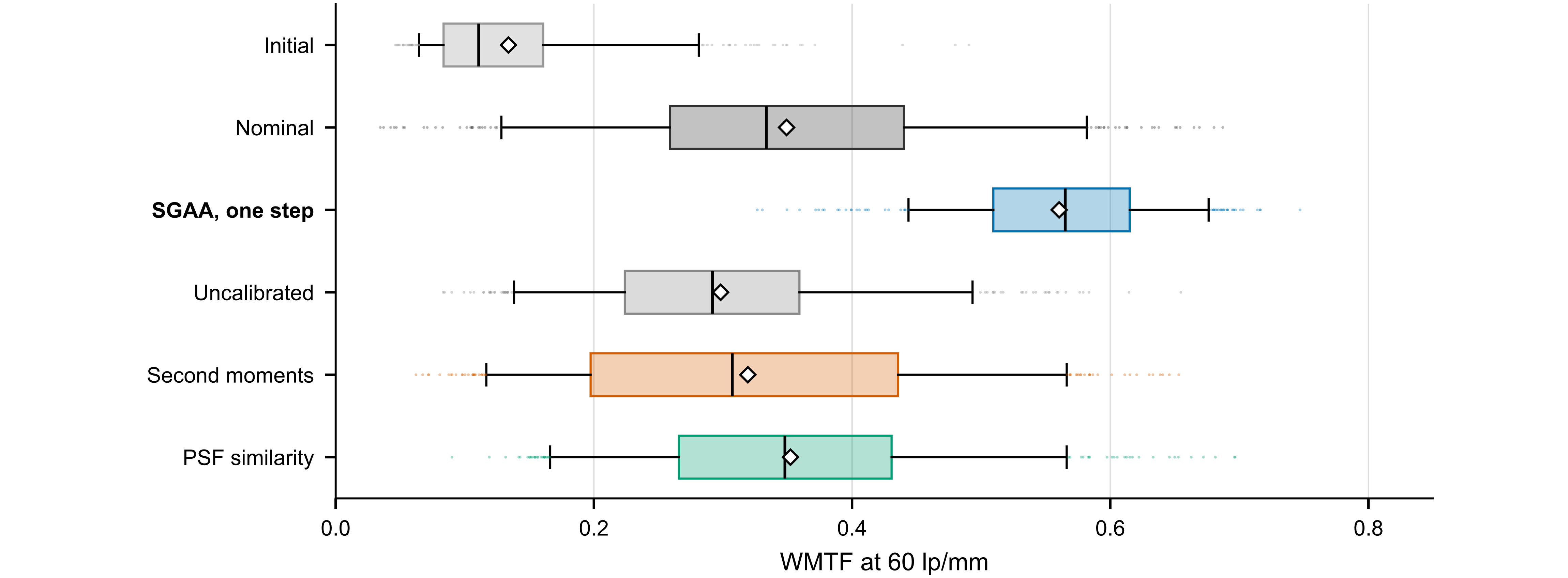}

\caption{Distributions of complete-lens WMTF60 after one-step AA using networks trained with different reference calibration schemes.}

\label{fig:simulation-calibration}

\end{figure}

The validation set used reference-corrected labels to evaluate regression fitting. The independent test set contained 18 specimens and used relative motor movements as ground truth. The prediction at each specimen's initial position was subtracted from its predictions at the sampled positions to evaluate the network's response to known relative movements. The mean absolute errors (MAEs) of SGAA for $\Delta x$, $\Delta y$, $\Delta r_x$, $\Delta r_y$, and $\Delta z$ were 1.585 $\mu$m, 1.399 $\mu$m, 0.594 arcmin, 0.690 arcmin, and 4.594 $\mu$m, respectively, as shown in Table~\ref{tab:calibration}.

\begin{table}[!htb]
\centering
\caption{Five-degree-of-freedom prediction MAEs and WMTF60 for the reference calibration schemes}
\label{tab:calibration}
\fontsize{8}{10}\selectfont
\setlength{\tabcolsep}{3pt}
\begin{tabular}{@{}lcccccc@{}}
\hline
Scheme & \shortstack{Δx\\(µm)} & \shortstack{Δy\\(µm)} & \shortstack{Δr\textsubscript{x}\\(arcmin)} & \shortstack{Δr\textsubscript{y}\\(arcmin)} & \shortstack{Δz\\(µm)} & \shortstack{WMTF60\\(n = 500)} \\
\hline
{\bfseries SGAA} & 1.585 & 1.399 & 0.594 & 0.690 & 4.594 & {\bfseries 0.5604 ± 0.0735} \\
Uncalibrated & 5.962 & 5.328 & 2.380 & 2.005 & 10.190 & 0.2982 $\pm$ 0.1083 \\
Second moments & 2.258 & 2.524 & 1.051 & 1.120 & 8.714 & 0.3193 $\pm$ 0.1433 \\
PSF similarity & 5.768 & 4.902 & 3.407 & 3.170 & 8.887 & 0.3523 $\pm$ 0.1197 \\
\hline
\end{tabular}
\par\smallskip
\begin{minipage}{\linewidth}\footnotesize
Note: MAEs use relative motor movements as ground truth. WMTF60 values are the one-step feedback results for the 500 alignment specimens.\par
\end{minipage}
\end{table}

Golden-sample reference calibration improves both relative-motion prediction and one-step alignment accuracy. As shown in Table~\ref{tab:calibration} and Fig.~\ref{fig:simulation-calibration}, SGAA had lower MAEs in all five degrees of freedom than the other three schemes, with substantially better network fitting performance than the other three schemes. For the 500 alignment specimens, the mean WMTF60 of SGAA was 0.5604, exceeding 0.2982 for \textit{Uncalibrated}, 0.3193 for \textit{Second moments}, and 0.3523 for \textit{PSF similarity}. SGAA obtained higher WMTF60 than \textit{PSF similarity} in 476/500 specimens, indicating that network-based reference calibration can guide the unified regressor toward alignment states with better image quality. Both results support the SGAA procedure: learning the misalignment response from local PSFs and using a common reference to guide calibration and subsequent unified regression.

\section{Experimental validation}\label{sec:experiments}

Following the five-degree-of-freedom validation using toleranced simulations in Section 3, this section evaluates SGAA for lateral pre-alignment of production G3 subgroups in an industrial setting. Physical alignment performance is assessed through recovery of the reference PSF morphology, cured subgroup states, complete-lens image quality, acceptance yield, and AA time.

\subsection{Setup, sampling, and network settings}\label{sec:experimental-setup}

\begin{figure}[!htb]
\centering
\includegraphics[width=\linewidth,height=0.43\textheight,keepaspectratio]{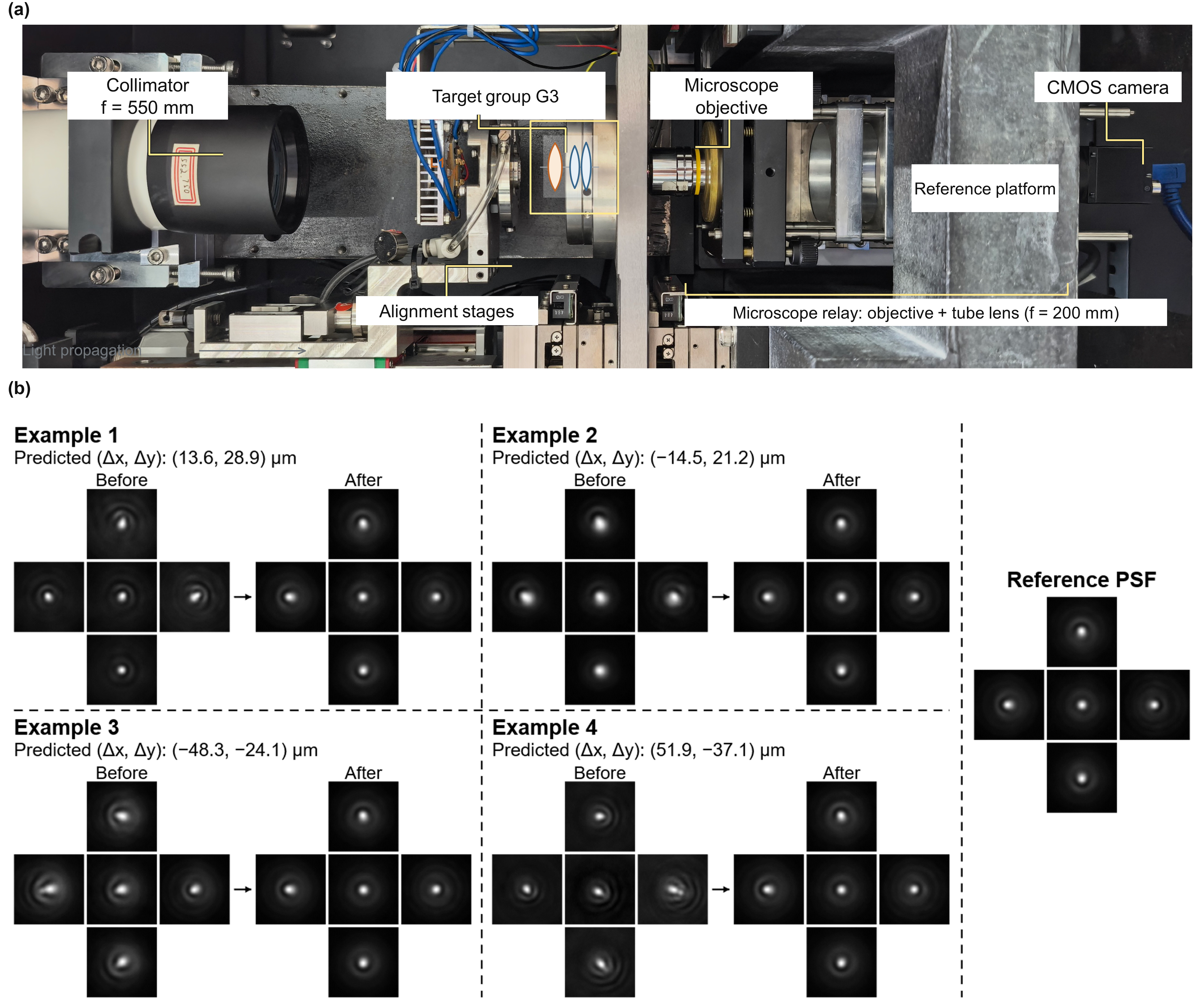}
\caption{Experimental setup and subgroup AA results. (a) Independent subgroup PSF observation setup. (b) Measured PSFs at five fields before and after AA for four subgroup examples and the network-predicted adjustments. Each example shows the center, right, left, top, and bottom fields; the reference PSFs are shown on the right.}
\label{fig:fig9}
\end{figure}

The experimental setup is shown in Fig.~\ref{fig:fig9}(a). A point source and collimator provide collimated illumination to G3. The spot formed by the subgroup at an intermediate image plane is relayed to a CMOS sensor using a 10$\times$ microscope objective with NA 0.3 and a tube lens. The main settings are listed in Table~\ref{tab:physical-settings}.

\begin{table}[!htb]
\centering
\caption{Optical and input settings for the physical experiments}
\label{tab:physical-settings}
\fontsize{8}{10}\selectfont
\setlength{\tabcolsep}{4pt}
\begin{tabular}{p{4.1cm}p{7.6cm}}
\hline
Item & Setting \\
\hline
Target & G3; adjustable L1, fixed L2 and L3 \\
Microscope objective & 10$\times$; NA = 0.3 \\
Collimator focal length & 550 mm \\
Tube-lens focal length & 200 mm \\
Object-side field height & (0,0), ($\pm$3,0), ($\pm$6,0), (0,$\pm$3), (0,$\pm$6) mm \\
Image input & Nine 128 $\times$ 128 grayscale PSFs; per-image min--max normalization \\
PSF image plane & Fixed \\
Lateral sampling range & $\Delta x$, $\Delta y$: $\pm$60 $\mu$m \\
Defocus sampling range & $\pm$60 $\mu$m \\
\hline
\end{tabular}
\end{table}

The experimental network uses a ResNet18 backbone. It follows the three stages described in Section 2: relative learning, golden-sample reference calibration, and unified regression. AdamW is used for optimization. Training over the three stages takes approximately 6 h on an RTX 5060 Ti GPU.

The acquired dataset contains approximately 30,000 sets of PSFs from 30 ordinary physical subgroup specimens. Both $\Delta x$ and $\Delta y$ are sampled over $\pm$60 $\mu$m, with defocus sampled over $\pm$60 $\mu$m to cover small departures from the image plane. Specimens are separated for unified regression: 21 are used for training and nine for validation. The 51 sets of PSFs acquired from one golden sample with small perturbations and defocus establish a common reference for the 30 physical subgroups with different tolerances and support unified-model training and validation.

The final validation MAEs are 0.804 $\mu$m for $\Delta x$ and 1.063 $\mu$m for $\Delta y$, and the two-dimensional RMSE is 1.677 $\mu$m. These values characterize how accurately the network fits the corrected labels. Alignment endpoints and the corresponding complete-lens image quality are evaluated in the subsequent experiments.

\subsection{Physical alignment and subgroup state}\label{sec:experimental-aa}

The 100 AA specimens were randomly sampled from one production batch, without overlap with training or validation specimens: 60 for SGAA and 40 for the Baseline. All specimens undergo decentering measurements and are assembled into complete lenses to image a resolution target for classification as 4 MP, 8 MP, or not accepted (NG), based on resolved line pairs. The Baseline in this section is an industrial dummy-lens AA scanning machine. Its compensating optical path has a configuration similar to the simulated dummy lens in Section 3: the remaining groups of the complete lens serve as the dummy lens, and G3 is aligned using through-focus MTF scanning.

During SGAA deployment, an unfixed G3 is first placed in the setup, and a suction nozzle holds its adjustable component L1. The network then predicts the AA correction and drives the motor. One focusing operation and fine AA adjustment are performed before adhesive dispensing and curing produce the aligned G3. A single deployment AA step takes approximately 0.4--0.5 s.

Figure~\ref{fig:fig9}(b) shows measured PSFs at five fields before and after AA for four subgroup examples and the network-predicted adjustments. In Examples 1, 3, and 4, the asymmetric rings visible before AA become less pronounced, and the multi-field spots more closely resemble the reference PSFs. Example 2 exhibits defocus associated with the specimen's mechanical structure, but its final PSFs also approach the reference. These images illustrate recovery of the PSF morphology toward the reference state in different specimens after SGAA.

\begin{table}[!htb]
\centering
\caption{Cured L1 positions, resolution grades, and mean AA times for the two methods}
\label{tab:physical-results}

\fontsize{8}{10}\selectfont
\setlength{\tabcolsep}{4pt}
\begin{tabular}{p{4.9cm}ll}
\hline
Metric & SGAA & Baseline \\
\hline
Records, n & 60 & 40 \\
X mean $\pm$ SD ($\mu$m) & 1.559 $\pm$ 7.360 & $-$4.910 $\pm$ 11.944 \\
Y mean $\pm$ SD ($\mu$m) & 0.153 $\pm$ 5.684 & $-$0.561 $\pm$ 8.693 \\
R median ($\mu$m) & 7.191 & 13.331 \\
8 MP, n (\%) & 49 (81.7) & 21 (52.5) \\
4 MP, n (\%) & 10 (16.7) & 13 (32.5) \\
NG, n (\%) & 1 (1.7) & 6 (15.0) \\
Acceptance yield & {\bfseries 81.7\%} & 52.5\% \\
Mean AA time (s) & {\bfseries 3.1} & 43 \\
\hline
\end{tabular}
\par\vspace{3pt}
\begin{minipage}{0.98\linewidth}\fontsize{8}{10}\selectfont Note: X and Y are the offsets of the cured L1 position from the geometric nominal position, measured using a reflection-type centering instrument relative to the center of the G3 optical axis, in $\mu$m. R is the magnitude of each specimen's two-dimensional decentering. Mean AA times exclude specimen loading, unloading, adhesive dispensing, and curing.\end{minipage}
\end{table}

Table~\ref{tab:physical-results} and Supplement 1, Fig. S2 summarize the cured L1 positions measured using a reflection-type centering instrument and the resolution grades for the 60 SGAA specimens and 40 Baseline specimens. SGAA yields a more concentrated distribution of aligned component positions, consistent with simulations indicating that tolerances in the compensating optical path may lead to larger subgroup alignment biases (Supplement 1, Section S3). Taking the 8 MP grade as the acceptance criterion, the acceptance yields are 81.7\% for SGAA and 52.5\% for the Baseline, an increase of 29.2 percentage points.

\subsection{Complete-lens image quality}\label{sec:experimental-quality}

\begin{figure}[!htb]
\centering
\includegraphics[width=0.96\linewidth,height=0.55\textheight,keepaspectratio]{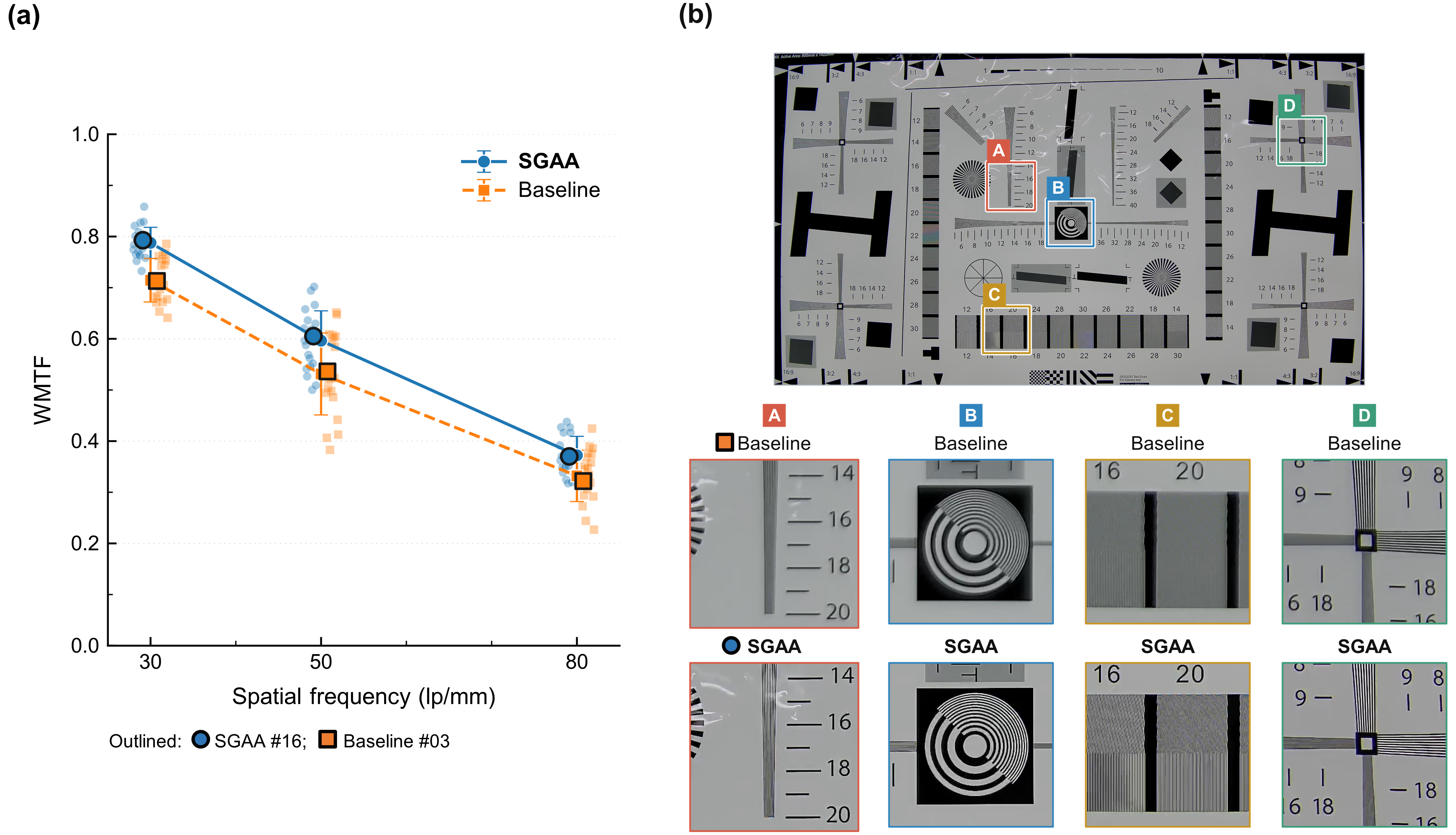}
\caption{Complete-lens image quality after SGAA and dummy-lens scanning. (a) WMTF distributions at the different frequencies. (b) Captured images of the test target using the complete lenses and comparisons of local details.}
\label{fig:fig11}
\end{figure}

Twenty G3 subgroups from each of the above groups are evaluated for MTF and image capture in the same complete-lens assembly. Figure~\ref{fig:fig11} compares the complete-lens image quality after subgroup alignment through frequency response and captured image details. The measured WMTF is the equally weighted mean of 10 MTF channels: tangential and sagittal responses at the center and four corners. As shown in Fig.~\ref{fig:fig11}(a), at 30, 50, and 80 lp/mm, the SGAA WMTF means $\pm$ sample standard deviations are 0.788 $\pm$ 0.030, 0.596 $\pm$ 0.058, and 0.372 $\pm$ 0.037, respectively. The means exceed the corresponding Baseline means by 0.073, 0.065, and 0.040. Thus, SGAA has higher mean WMTF and smaller between-specimen variation at all three frequencies in this dataset. Figure~\ref{fig:fig11}(b) compares the complete-lens images for SGAA Example 16 and Baseline Example 3. In the SGAA example, the wedge patterns, ring target, and fine lines are more readily resolved, and the edges of the cross targets are clearer.

Notably, in through-focus MTF measurements, SGAA reduced the mean range between the maximum and minimum multi-field peak positions by 44.5\% (Supplement 1, Section S3 and Fig. S1). Together with the multi-frequency WMTF and captured images, these results show that subgroups pre-aligned using SGAA provide higher resolving quality, smaller between-specimen variation, and more closely grouped multi-field focus positions in the complete lens than the industrial Baseline.

\FloatBarrier
\Needspace{8\baselineskip}
\section{Conclusions and discussion}\label{sec:conclusions}

We propose SGAA, an independent subgroup active-alignment method for the sequential assembly of multi-group lenses. Golden-sample reference calibration and multi-field PSF learning convert misalignment responses under aberrated imaging conditions into component adjustments toward a common optical reference. In simulations with tolerances, one feedback step reduced the defocus-removed differential RMS by 85.4\% and increased the complete-lens WMTF60 from 0.1338 to 0.5604. In physical AA, the 8 MP acceptance yield increased by 29.2 percentage points relative to industrial dummy-lens scanning, the mean AA time decreased from 43 s to 3.1 s, and the mean complete-lens WMTF increased at all three evaluated frequencies. SGAA thus translates restoration of the subgroup reference state into gains in complete-lens resolving quality and assembly efficiency.

This study has validated single-subgroup alignment in two lens designs.

Future work will apply single-step pre-alignment separately to multiple groups within the same lens and link each group's local adjustment to complete-lens evaluation, establishing a closed-loop assembly process from components through subgroups to the complete lens. We will further examine reference transfer and coordinated adjustment across different lens structures and assembly conditions, determine observation configurations and sampling ranges from PSF responses to misalignment, and extend subgroup alignment to different optical architectures, providing a research foundation for broader industrial applications of the method.

\section*{Acknowledgments}

The authors thank Dongguan Yutong Optical Technology Co., Ltd. for providing lens samples and support.

\section*{Disclosures}

The authors declare no conflicts of interest.

\section*{Data availability}

The industrial experimental data are not publicly available due to confidentiality restrictions. Access may be granted upon reasonable request to the corresponding author, subject to approval by the data owner.

\clearpage
\setcounter{section}{0}
\setcounter{subsection}{0}
\setcounter{equation}{0}
\setcounter{figure}{0}
\setcounter{table}{0}
\renewcommand{\thesection}{S\arabic{section}}
\renewcommand{\theequation}{S\arabic{equation}}
\renewcommand{\thefigure}{S\arabic{figure}}
\renewcommand{\thetable}{S\arabic{table}}
\renewcommand{\theHsection}{supp.\arabic{section}}
\renewcommand{\theHequation}{supp.\arabic{equation}}
\renewcommand{\theHfigure}{supp.\arabic{figure}}
\renewcommand{\theHtable}{supp.\arabic{table}}
\setlength{\textfloatsep}{9pt plus 2pt minus 2pt}
\setlength{\intextsep}{9pt plus 2pt minus 2pt}
\title{SGAA: Independent active alignment of optical subgroups with uncompensated aberrations\\Supplemental document}

\section{Supplementary derivations}
Building on the reference-field recovery relation in Eq.~(9) of the main text, this section uses incoherent intensity summation and the Fourier relation between the PSF and MTF to explain how recovery of the subgroup reference state translates into recovery of complete-lens image quality.

Under identical illumination and evaluation conditions, let $V_{h,\lambda}(q)$ denote the complex field at the complete-lens image plane, where $h$ and $\lambda$ denote the field of view and wavelength, respectively, and $r$ is the image-plane coordinate. Intensities at different wavelengths are summed incoherently with spectral weights $w_\lambda$ and normalized by total energy to obtain the PSF. The magnitude of its Fourier transform gives the MTF:
\begin{equation}
\begin{gathered}
J_h(q)=\sum_\lambda w_\lambda|V_{h,\lambda}(q)|^2,\\
p_h(q)=\frac{J_h(q)}{\int J_h(q)\,\mathrm dr},\qquad
\operatorname{MTF}_h(q)=\left|\mathcal F[p_h(q)]\right|.
\end{gathered}\label{eq:S1}
\end{equation}
With fixed spectral weights and nonzero reference energy, $L^2$ convergence of the complex field yields $L^1$ convergence of the intensity and normalized PSF, thereby recovering the MTF. Recovery of the MTF at all fields, frequencies, and tangential/sagittal components also restores their fixed, nonnegative, normalized weighted sum, WMTF.

The reference state includes manufacturing tolerances, so its image quality is somewhat degraded relative to the tolerance-free design. By recovering the optical response of this reference state, SGAA compensates for the additional aberrations introduced by component misalignment and restores complete-lens image quality to this practical reference level.

\section{Supplementary experimental and algorithm details}
This section supplements the three-stage training and feedback procedure in the main text with image preprocessing, supervision objectives, and deployment settings.

\textbf{Preprocessing.} The simulation workflow is: toleranced G3 $\rightarrow$ five-degree-of-freedom sampling $\rightarrow$ nine-field PSFs $\rightarrow$ individual-field image processing $\rightarrow$ offline noise addition $\rightarrow$ nine-channel network input. The quantization steps are $0.2~\mu$m for $\Delta x$, $\Delta y$, and $\Delta z$, and $0.2'$ for $\Delta r_x$ and $\Delta r_y$. Both simulated and measured PSFs are cropped about their centroids to $128\times128$ grayscale images. Measurements use a 1-inch industrial camera with $4080\times3072$ pixels.

Offline perturbation applies subpixel deformation followed by grayscale noise to simulate real alignment conditions. The X/Y displacement components of each pixel are independently drawn from zero-mean Gaussian distributions with a standard deviation of 0.4 pixels. The grayscale noise is zero-mean Gaussian noise with an SNR of 28 dB and standard deviation $\sigma_n=\sqrt{\operatorname{mean}(I_{\mathrm{warp}}^2)/10^{2.8}}$, where $I_{\mathrm{warp}}$ is the deformed image. Each field image is min--max normalized, and the images are stacked into a $9\times128\times128$ input. Measured input images undergo simple smoothing.

Coordinate residuals are normalized by the scale of each degree of freedom to accommodate the different units of translation and tilt. The five simulation scales, in the order $\Delta x$, $\Delta y$, $\Delta r_x$, $\Delta r_y$, and $\Delta z$, are $(60~\mu\mathrm m,60~\mu\mathrm m,20~\mathrm{arcmin},20~\mathrm{arcmin},60~\mu\mathrm m)$. In the measurements, the sampling ranges of $\Delta x$, $\Delta y$, and defocus are all $\pm60~\mu$m.

\textbf{Stage 1: relative pretraining.} A shared feature extractor and specimen-specific regression branches learn the correspondence between PSF changes and acquisition-motion differences. The relative term uses smooth L1 loss ($\beta=0.5$), while absolute golden-sample labels anchor the reference and blur-consistency regularization enforces consistent predictions. The three terms have weights of 1, 0.5, and 0.2, respectively.

\begin{table}[!htbp]
\centering
\caption*{Table S1. Layer-by-layer architecture of the calibration network.}
\small
\setlength{\tabcolsep}{4pt}
\renewcommand{\arraystretch}{1.10}
\begin{tabularx}{\linewidth}{@{}l c r c >{\raggedright\arraybackslash}X@{}}
\hline
\textbf{Layer (type)} & \textbf{Input shape} & \textbf{Input elements} & \textbf{Output shape} & \textbf{Description}\\
\hline
Conv2d + GN + ReLU & $(9,128,128)$ & 147456 & $(64,64,64)$ & $7\times7$, stride 2\\
MaxPool2d & $(64,64,64)$ & 262144 & $(64,32,32)$ & $3\times3$, stride 2\\
Residual stage 1 & $(64,32,32)$ & 65536 & $(64,32,32)$ & Stride 1\\
Residual stage 2 & $(64,32,32)$ & 65536 & $(128,16,16)$ & First block: stride 2\\
Residual stage 3 & $(128,16,16)$ & 32768 & $(256,8,8)$ & First block: stride 2\\
Residual stage 4 & $(256,8,8)$ & 16384 & $(512,4,4)$ & First block: stride 2\\
AdaptiveAvgPool2d & $(512,4,4)$ & 8192 & $(512,1,1)$ & Global average pooling\\
Flatten & $(512,1,1)$ & 512 & $(512)$ & Flattening\\
FC & $(512)$ & 512 & $(16)$ & Shared features\\
Affine & $(16)$ & 16 & $(d)$ & Specimen/golden branches\\
\hline
\end{tabularx}
\end{table}\FloatBarrier
Each residual stage contains two BasicBlocks; convolutional normalization uses GN.

\textbf{Stage 2: reference calibration.} The Stage 1 model is inherited to fit local acquisition coordinates while introducing a common-reference constraint. Table S1 gives the calibration network architecture. Let the specimen-specific branch prediction be $p_i(y)=h_i(E_\theta(y))$; the golden sample uses a separate regression branch. With $i$, $j$, and $g$ indexing ordinary specimens, acquisition states, and golden-sample observations, respectively, the three absolute-supervision terms are
\begin{equation}
\begin{aligned}
L_g&=\mathbb E_g\,\ell_{\mathrm{abs}}\!\left(p_{\mathrm{gold}}(y_g),u_g^{\mathrm{ref}}\right),\\
L_{\mathrm{local}}&=\mathbb E_{i,j}\,\ell_{\mathrm{abs}}\!\left(p_i(y_{ij}),u_{ij}\right),\\
L_{\mathrm{ref}}&=\ell_{\mathrm{abs}}\!\left(\mathbb E_{i,g}[p_i(y_g)-u_g^{\mathrm{ref}}],0\right).
\end{aligned}\label{eq:S2}
\end{equation}
Here, $\ell_{\mathrm{abs}}$ is the component-wise mean square of the normalized residual. $L_g$ and $L_{\mathrm{local}}$ supervise the golden-sample labels and local coordinates, respectively. $L_{\mathrm{ref}}$ averages the signed residuals over ordinary-specimen regression branches and golden-sample observations before squaring, thereby constraining the population mean offset. The weights of these three terms and the relative loss are 0.1, 0.8, 0.01, and 0.2, respectively.

The zero-coordinate PSF from the golden-sample training set serves as a fixed optical reference to further constrain the image response near the estimated center:
\begin{equation}
\begin{aligned}
d_{ijh}&=\alpha\frac{1-\operatorname{SSIM}(I_{ijh},I_{\mathrm{ref},h})}{2}
+(1-\alpha)\frac{1}{2}\left\|P_{ijh}-P_{\mathrm{ref},h}\right\|_1,\\
d_{ij}&=\frac{1}{H}\sum_h d_{ijh},\\
L_{\mathrm{PSF}}&=\frac{1}{N}\sum_i\sum_{j\in\mathcal N_K(\widehat O_i)}\omega_{ij}(\widehat O_i)d_{ij}.
\end{aligned}\label{eq:S3}
\end{equation}
Here, $I$ denotes nonnegative intensity, and $P$ is the energy-normalized distribution. Each image is normalized by its peak before SSIM comparison. The neighborhood consists of the $K$ states nearest to the estimated center $\widehat O_i$, and $\omega$ is the Gaussian distance weight normalized within the neighborhood ($\sigma=0.2$).

\textbf{Stage 3: unified regression.} Labels for ordinary specimens are corrected using the reference-offset estimates in Eq.~(7) of the main text. With these offsets fixed, the unified regression network is initialized with ImageNet weights and trained using the squared L2 coordinate error per sample. All three stages use AdamW with a batch size of 16. The three simulation stages were trained for 10, 53, and 27 epochs, respectively, with model selection based on the validation set.

\textbf{Algorithm deployment.} Multi-field PSFs acquired at a fixed image plane undergo the same preprocessing and are input to the unified network to predict misalignment relative to the reference. The output is converted directly into stepper-motor commands. After L1 is clamped, PSF acquisition, network prediction, and motor adjustment form the feedback loop, together with one focusing step and fine adjustment. The procedure stops when the network prediction error is below $1.5~\mu$m, typically converges in 2--4 iterations, and is followed by adhesive curing. The mean AA time is 3.1 s.

The comparison scan follows Ref.~\cite{suppLiu2024}, comparing WMTF axis by axis from coarse to fine, reducing the step size, and retaining the best position, without additional position bounds. The scan budget is capped at 50 scans, whereas \textit{full} has no budget cutoff. Final complete-lens evaluation finely adjusts the image plane to locate the best image-quality position and calculate WMTF.

\section{Supplementary experimental data}
\begin{figure}[!htbp]
\centering
\captionsetup{width=\linewidth}
\includegraphics[width=0.85\linewidth]{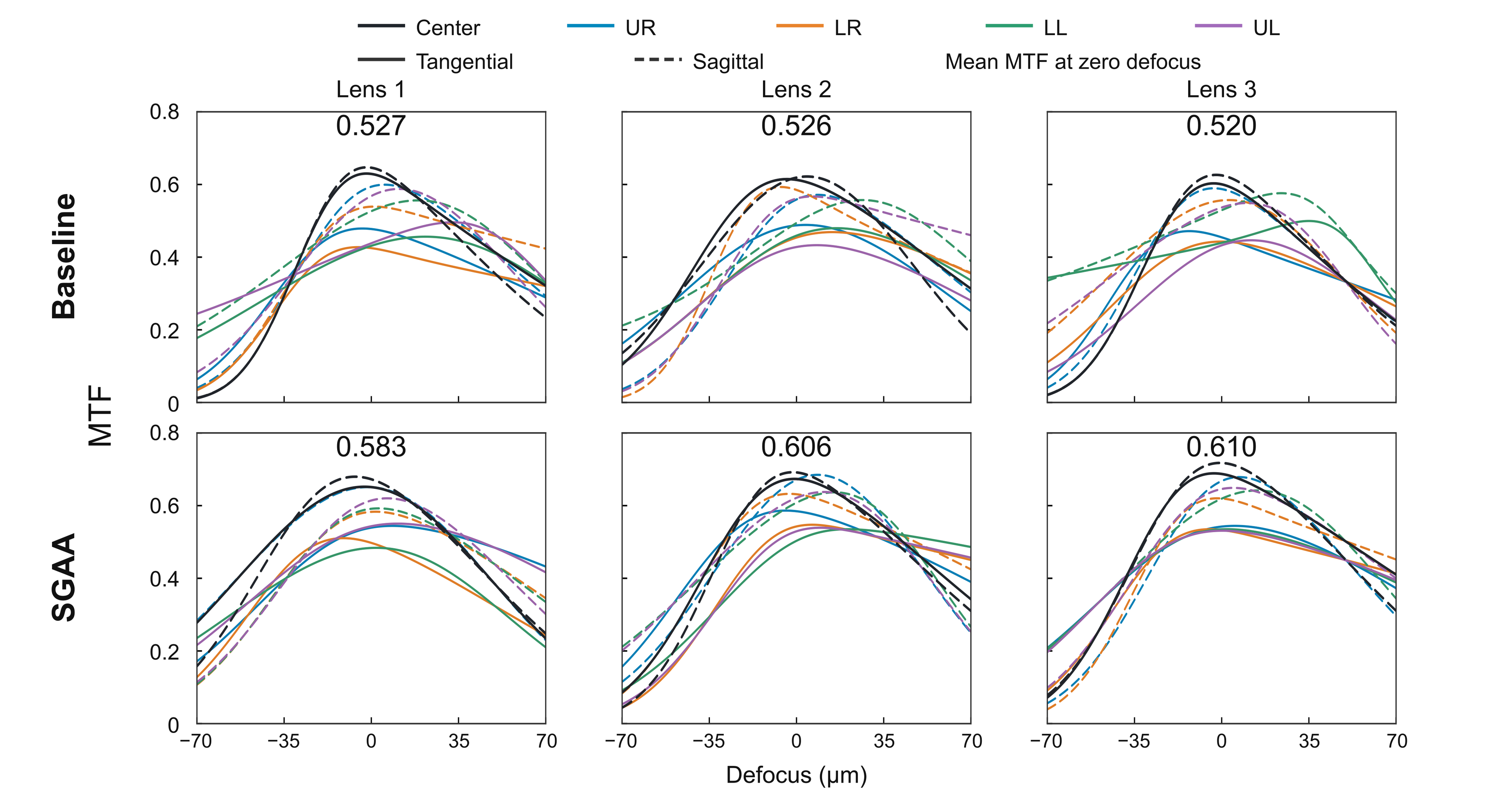}
\caption*{Fig. S1. Complete-lens through-focus MTF after alignment by the two methods (50 lp/mm).}
\end{figure}\FloatBarrier
Figure S1 compares the complete-lens through-focus responses of the same three G3 specimens after Baseline and SGAA alignment; each column corresponds to the same specimen. Colors distinguish the central field and four off-axis fields, while solid and dashed curves indicate tangential and sagittal responses, respectively. The numbers in the figure are the equal-weight means of the ten channels across five fields at zero defocus.

SGAA improves both image contrast and the consistency of best-focus positions across channels. The average Mean MTF at zero defocus increases from 0.525 to 0.599, and the peaks of all ten channels exceed the corresponding Baseline values. The mean range of peak positions decreases from $39.583~\mu$m to $21.950~\mu$m, a reduction of 44.5\%, indicating closer agreement among the best-focus positions of different fields and tangential/sagittal responses. These comparisons demonstrate the contribution of local subgroup alignment to complete-lens imaging performance in terms of both MTF level and focal-plane consistency.

\begin{figure}[!htbp]
\centering
\captionsetup{width=\linewidth}
\includegraphics[width=0.53\linewidth]{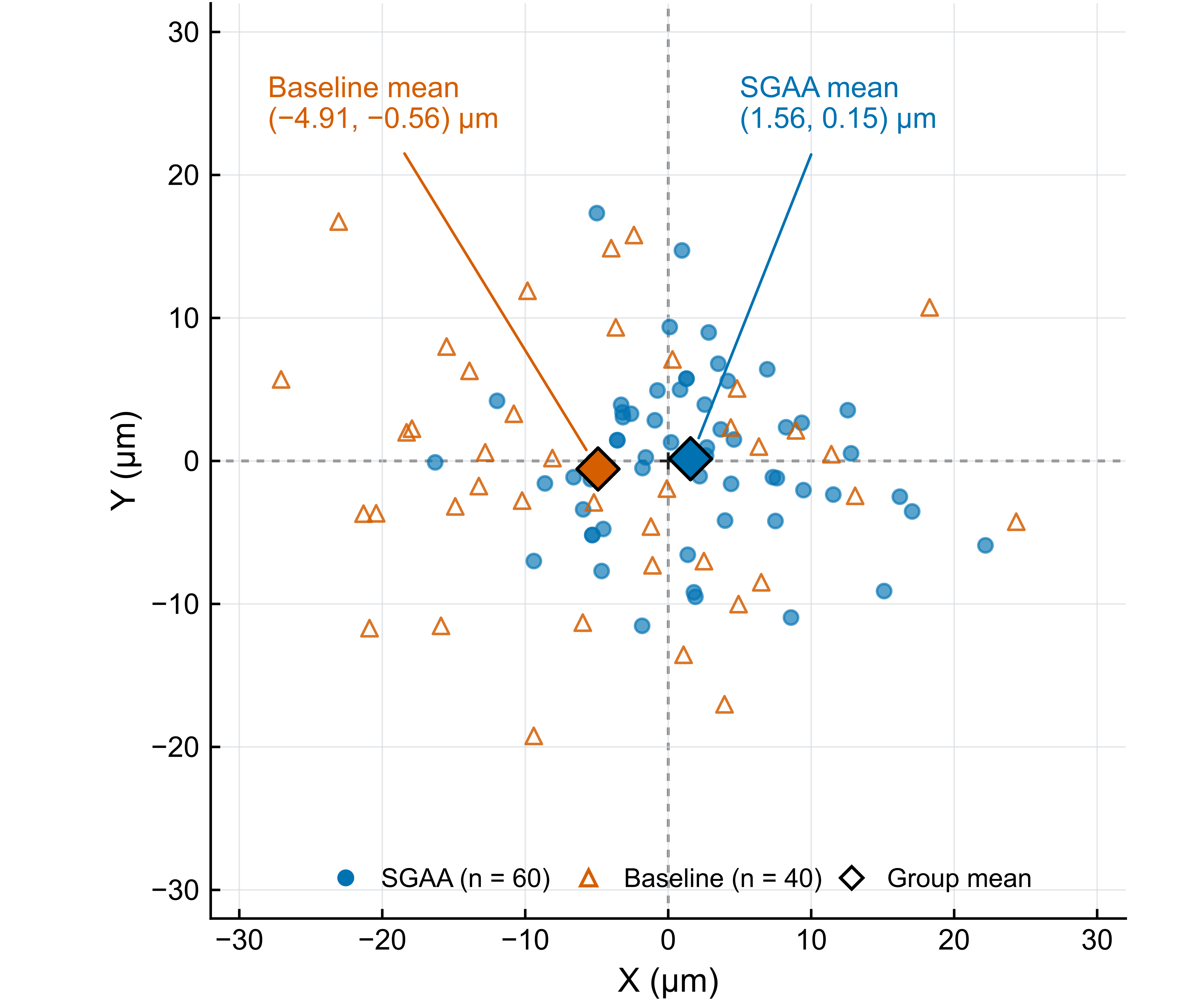}
\caption*{Fig. S2. Two-dimensional decenter distribution of cured L1 positions after alignment by the two methods.}
\end{figure}\FloatBarrier
Figure S2 shows the two-dimensional decenter of L1 within the subgroup after curing relative to its geometric nominal position, measured with a reflection-type centering instrument using the G3 optical axis as the reference. Blue circles and orange triangles represent the 60 SGAA specimens and 40 Baseline specimens, respectively; diamonds indicate group means.

The SGAA distribution is centered closer to the geometric nominal zero and is more concentrated, consistent with simulations indicating that tolerances in the compensating optical path may lead to larger subgroup alignment biases. The mean $(\Delta x,\Delta y)$ values of the two groups are $(1.559,0.153)~\mu$m and $(-4.910,-0.561)~\mu$m, respectively. The sample standard deviations along the two directions are 7.360 and $5.684~\mu$m for SGAA, both below the Baseline values of 11.944 and $8.693~\mu$m. These position measurements support SGAA's ability to align components within a subgroup and corroborate the 8 MP acceptance yields of 81.7\% versus 52.5\% for the same batches reported in the main text, showing that SGAA achieves both a concentrated alignment-position distribution and a higher complete-lens imaging yield.

\end{document}